\documentclass{aa}  

\usepackage{graphicx}

\usepackage{txfonts}
\usepackage{newtxtext}
\usepackage[varvw]{newtxmath}

\begin{document}

   \title{Active asteroid P/2023 JN16: An impulsive event ejected millimetre-sized debris without excavating water ice}

   \subtitle{}

   \author{J. Agarwal\inst{1}   }

   \institute{Institute of Geophysics and Extraterrestrial Physics, TU Braunschweig, Mendelssohnstr. 3, D-38106 Braunschweig\\
     \email{j.agarwal@tu-braunschweig.de}
}

   \date{}
  \abstract
   {Quantifying the efficiency of impacts and rotational disintegration of asteroids to excavate ice will ultimately improve our understanding of the abundance and depth of water ice in these objects.}
   {We searched for perihelion re-activation of asteroid P/2023 JN16 (Lemmon) that had previously undergone an impulsive event of dust emission far from perihelion. We also studied the dust sizes, ejection velocities, and colours, with an aim to constrain the nature of the dust emission event.}
   {We analysed images obtained with ESO's VLT/FORS2 instrument and from the ZTF archive.}
   {We do not find any signs of re-activation at perihelion, constraining the depth to ice in this asteroid to at least 35\,cm, but compatible also with tens of metres. The debris is characterised by a peculiar absence of small, fast-dissipating dust, arguing against an impact origin for the dust. Most of the scattering cross-section is given in particles of radius $>$3mm. The ejection velocity significantly influences the orientation of the debris trail, such that a synchrone-derived emission date would have been misleading.}
   {}

   \maketitle
   \nolinenumbers

\section{Introduction}
\label{sec:intro}

Comets originate from two reservoirs of small bodies in the outer Solar System, the scattered disc in the transneptunian region and the Oort Cloud. Cometary orbits have significant eccentricities and are therefore subject to frequent change due to the gravitational interaction with planets. In contrast, main-belt asteroids orbit the Sun on rather stable, low-eccentricity orbits between those of Mars and Jupiter. The value of a small body's Tisserand parameter with respect to Jupiter ($T_J$) indicates whether ($T_J < 3$) or not ($T_J > 3$) its orbit intersects the orbit of Jupiter.

Temperatures in the asteroid belt are too high for surface water ice or other typical cometary volatiles to be stable on Solar System timescales \citep{schorghofer2008,schorghofer2016,schorghofer-hsieh2018, prialnik-rosenberg2009, capria-marchi2012}. Therefore, most asteroids do not regularly emit material to interplanetary space, which distinguishes them from comets. When, by exception, an asteroid ($T_J > 3$) is observed to have emitted dust, it is called an `active asteroid' \citep[e.g. ][]{jewitt2012, jewitt_AIV}.
Possible causes of dust emission from asteroids include collisions \citep[e.g.][]{ishiguro-hanayama2011a, ishiguro-hanayama2011b, kim-ishiguro2017a,kim-ishiguro2017b, jewitt-kim2023_dimorphos, kim-jewitt2023_dimorphos} and mass shedding or disintegration by fast rotation \citep[e.g. ][]{drahus-waniak2015, jewitt-li2021}.

The main-belt comets \citep[MBCs;][]{hsieh-jewitt2006} are a sub-group of the active asteroids that have shown recurrent activity during at least two perihelion passages, indicating the action of a temperature-sensitive process such as water ice sublimation. Water vapour in the MBCs 238P, 358P and 133P has now been confirmed spectroscopically with the James Webb Space Telescope \citep[JWST, ][]{kelley-hsieh2023, hsieh-noonan2025, hsieh-noonan2026}. The presence of water ice has also been inferred from surface spectra of 24 Themis \citep{rivkin-emery2010, campins-hargrove2010} and from the detection of water vapour near Ceres \citep{kueppers-orourke2014}.

Given the short lifetime of surface water ice in the asteroid belt, the water ice driving the dust emission from MBCs is thought to have been recently excavated; for instance, by a collision or structural change induced by fast rotation. However, most of the known MBCs were discovered when already sublimating. The dust emission from other asteroids has convincingly been attributed to `impulsive' events, but many of these objects did not re-activate at perihelion. In this paper, we report on our search for reactivation of asteroid P/2023 JN16 (JN16, henceforth) that emitted dust in spring 2023 \citep{CBET_JN16, deen_mpec_JN16_aug24} and went through perihelion in December 2024.

\begin{figure}
  \resizebox{\hsize}{!}{\includegraphics{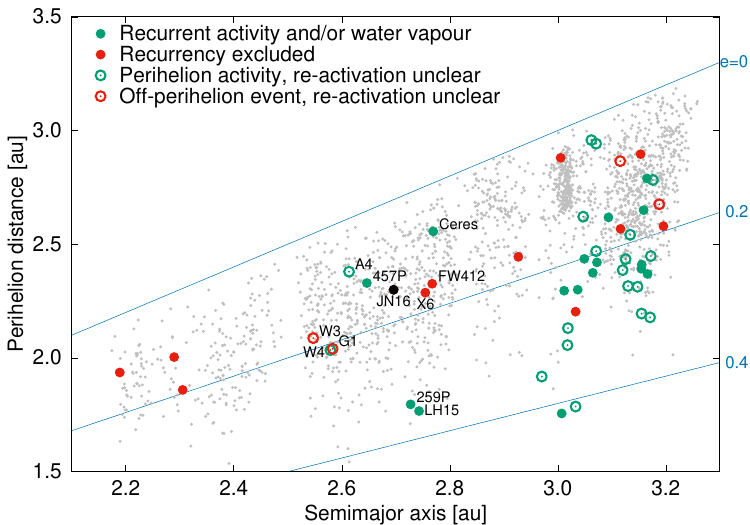}}
  \caption{Perihelion distance (as a direct measure of maximum solar heating) versus semi-major axis of small bodies at asteroid belt distances. Grey dots represent  asteroids with $H\!\!\!<$12 reported by the Minor Planet Center on 01 October 2025 (https://www.minorplanetcenter.net/data). Green and red symbols represent the known active asteroids, and the black point indicates the orbital location of P/2023 JN16. Middle-belt objects are labelled for cross-reference with Table~\ref{tab:middlebelt}, where also their full designations are listed. Blue lines indicate orbits of constant eccentricity, $e$, as labelled on the righthand margin.}
  \label{fig:aa_aq}
\end{figure}

Figure~\ref{fig:aa_aq} shows the orbital properties of the active main belt asteroid population, as known to the author in November 2026. Not counting JN16, there are 16 MBCs that have shown recurrent activation at perihelion. In addition, 11 objects have impulsively emitted dust and debris (two of them did so multiple times) without re-activating at perihelion. For 23 asteroids, the data are not sufficient to determine whether or not their activity is recurrent. Of these, 19 have been observed active during a single perihelion passage and 4 emitted dust far from perihelion. 

The asteroid JN16 was discovered on 10 May 2023 in images from the Mt. Lemmon Survey's 1.5-m reflector, having the appearance of an asteroidal (i.e. point source-like) object \citep{CBET_JN16}. Its orbital elements (semi-major axis, $a$=2.696\,au, eccentricity, $e$=0.147, and inclination with respect to the ecliptic, $i$=3.7$^\circ$) place it in the middle main-belt (Fig.~\ref{fig:aa_aq}). At the time of discovery, JN16 was at a true anomaly of $\nu$=216$^\circ$, heliocentric and geocentric distances of $r_h$=2.996\,au and $\Delta$=2.003\,au, and phase angle of $\alpha$=4.35$^\circ$.  

\begin{figure}
  \resizebox{\hsize}{!}{\includegraphics{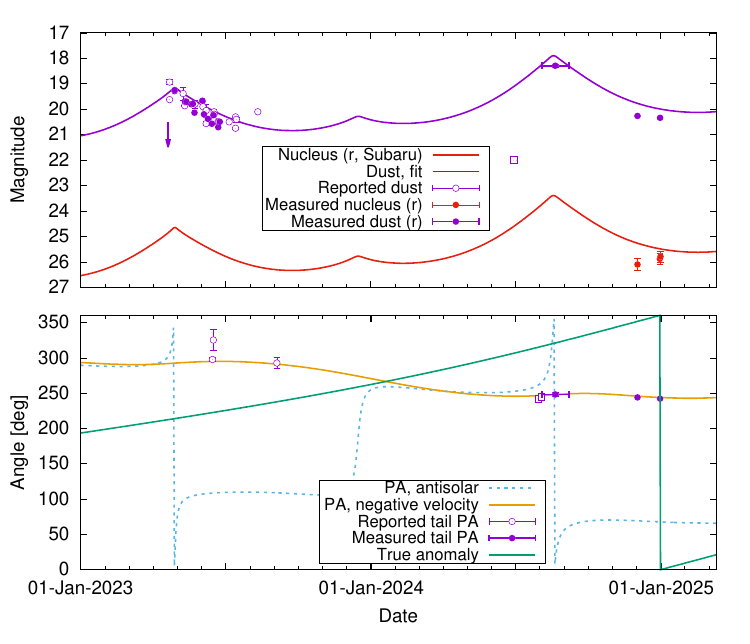}}
  \caption{Measurements reported in \citet{CBET_JN16} and \citet{deen_mpec_JN16_aug24} (open symbols), and geometric context, as well as our photometric measurements derived in Sections~\ref{sec:nucleus} -- \ref{sec:tail} (filled symbols). In the upper panel, the red line indicates the expected apparent r-band magnitude of the bare, pre-event nucleus based on the 2018 Subaru measurement (see text). The reported apparent magnitudes of the dust are not corrected for filter, while our measurements are in the r band. The square symbol represents a measurement with unknown aperture size, and the vertical arrow indicates the upper limit derived from ZTF for 27 April 2023. The violet line indicates the apparent magnitude corresponding to an object of absolute magnitude $H_R$=15 and a phase function parameter $G$=0.2. This is a visual fit to the data. In the bottom panel, violet symbols indicate the PAs of the tail. The two square symbols correspond to reported PAs of 62$^\circ$ and 65$^\circ$, after adding 180$^\circ$ (see text). The horizontal error bar indicates the observing time interval of stacked ZTF images (Sec.~\ref{sec:ZTF_2024}).}
  \label{fig:reported_data}
\end{figure}

\citet{CBET_JN16} found JN16 in archival pre-discovery images obtained with the  8.3-m Subaru Telescope at Mauna Kea on 18.5 March 2018. They reported a stellar appearance and an i-band apparent magnitude of $M_{\rm i} = 24.7 \pm 0.1$. This measurement likely shows the pre-event, bare nucleus. We converted this apparent magnitude, $M_{\rm app}$, to an absolute magnitude, $H$, using
\begin{equation}
  H = M_{\rm app} - 5\log_{10}(r_h \Delta) + 2.5\log_{10}(\Phi(G,\alpha)),
  \label{eq:H}
  \end{equation}
where $r_{\rm h}$, $\Delta$ and $\alpha$ are the heliocentric and geocentric distances and the phase angle at the time of observation (values are listed in Table~\ref{tab:vlt_obs}). $\Phi(G,\alpha)$ is the phase function according to the HG-formalism \citep{bowell-hapke1989}. Due to the unknown spectral type of JN16, we assumed $G = 0.20 \pm 0.05$, such that $H_{\rm i, pre} = 20.51 \pm 0.11$. For comparison with other data, we converted this to r-band $H_{\rm r, pre} = 20.63 \pm 0.11$ using solar colour ($r-i$) = 0.12 \citep{willmer2018}. The corresponding apparent r-band magnitude as a function of time is shown as the solid red line in the upper panel of Figure~\ref{fig:reported_data} and will be further discussed in Section~\ref{sec:nucleus} of this paper.

Photometry carried out throughout spring 2023 (reported in \citet{CBET_JN16} and \citet{deen_mpec_JN16_aug24} and shown by open symbols in Fig.~\ref{fig:reported_data}) yielded absolute magnitudes about 5.5\,mag brighter than this. This suggests that JN16 was surrounded by a cloud of dust scattering sunlight, supported by the observation that its appearance was less compact than that of neighbouring stars.

The reported photometry on r-band images from the Zwicky Transient Facility (ZTF) yielded an upper brightness limit of r$>$20.5 (apparent magnitude) for UT 27 April 2023, 08:17, and a detection at r = 19.1 on UT 29 April 2023, 06:57, which suggests that some impulsive event caused JN16 to shed dust and brighten by at least 1.4\,mag during this 2-day interval \citep{CBET_JN16}, at true anomaly 214$^\circ$. This nearly coincides with the object's opposition at $\alpha$=0.8412$^\circ$ on UT 29 April 2023, 12:19. Fig.~\ref{fig:reported_data} shows that the intrinsic brightness of the object remained unchanged until the end of July 2023, as noted by \citet{CBET_JN16}. The presence of a tail begins to be reported as of June 2023. The first two reports mention a tail length of (1-2)\arcsec, barely longer than a typical seeing disc, which probably affects the accuracy of the position angle (PA, measured counterclockwise from north) measurement. All subsequently reported PAs (bottom panel of Fig.~\ref{fig:reported_data}) are consistent with the projected negative heliocentric velocity vector (yellow line), where large and old particles are found, while freshly emitted, small particles should have been located near the projected heliocentric position vector (dashed blue line). Two of the reported PAs are from data sets that do not include a nucleus detection, hence the reported tail orientation carries an ambiguity of 180$^\circ$. We plot them in Fig.~\ref{fig:reported_data}\,(bottom) with square symbols after adding 180$^\circ$.

In this paper, we re-visit the archival ZTF data from 2023 along with  ZTF images from 2024 that had not been analysed previously. We also present new images obtained with the FOcal Reducer and low dispersion Spectrograph (FORS2) of the European Southern Observatory's (ESO's) 8m Very Large Telescope (VLT) around the time of JN16's perihelion passage in December 2024. Section~\ref{sec:obs} describes the observations and data processing. Measured quantities and derived dust and nucleus properties are presented in Section~\ref{sec:results}. These findings are discussed in the wider context of asteroid activity in Section~\ref{sec:discussion} and summarised in Section~\ref{sec:conclusion}.

\section{Observations and data processing}
\label{sec:obs}

\subsection{Archival 2023 ZTF images}
We downloaded from the ZTF archive\footnote{https://irsa.ipac.caltech.edu/applications/ztf} all images of JN16 obtained between 01 January 2023 and 30 June 2023 in snippets of 30\arcsec\ $\times$ 30\arcsec\ (31$\times$31 pixels). These images were taken in ZTF r-, g-, or i-bands with exposure times of 30\,s. A bright source of stellar appearance, but no tail, is visible at the position of JN16 in all images obtained between 29 April and 25 May (Table~\ref{tab:ztf}). There are no images from 26 May to 02 June. As of 03 June 2023, no source can be visually discerned at the predicted position of JN16. The same applies to all 2023 ZTF images obtained on or before 27 April, as reported by \citet{CBET_JN16}. 

We measured the flux in circular apertures of 2\,px radii and subtracted the background approximated by the median in an annulus having inner and outer radii of 4 \,px and 7\,px using the Image Reduction and Analysis Facility (IRAF)\footnote{https://iraf.noirlab.edu/}. We employed the zeropoints for PSF photometry (header keyword MAGZP, in the AB system) and aperture correction terms for aperture diameter of 4\,px (keyword APCOR3) to calculate the apparent magnitude, $M_{\rm app}$, of JN16 (Table~\ref{tab:ztf}). For the data from 27 April, we derived upper limits by adding the scaled image of a star to the images of 27 April, reducing the scaling factor to the point where the star could not any more be visually distinguished from the background noise. Our findings are consistent with \citet{CBET_JN16} for the same data (Fig.~\ref{fig:reported_data}).

\begin{figure}
  \resizebox{\hsize}{!}{\includegraphics{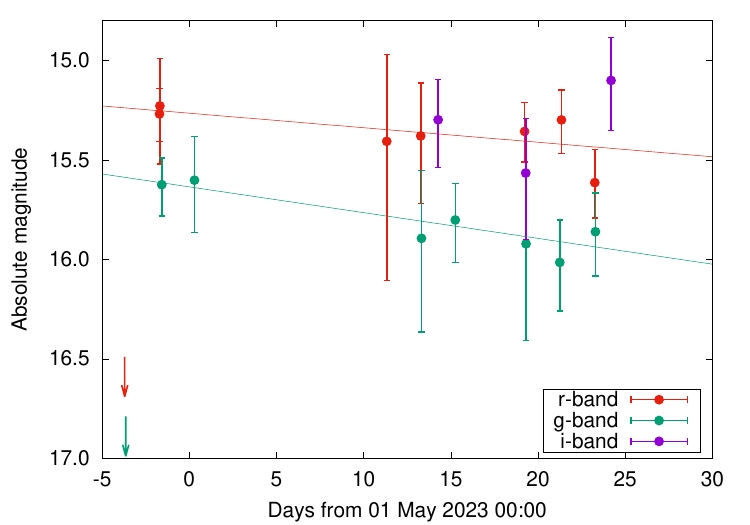}}
  \caption{Absolute magnitude (in AB system) of JN16 in ZTF images from late April and May 2023. The lines represent linear fits to the data, with slopes of 0.0073\,mag/day in the r band and 0.0130\,mag/day in the g band. The error bars mainly reflect the statistical noise of the background but also include the uncertainty resulting from the unknown phase function (see text). Arrows represent upper limits.}
  \label{fig:ztf}
\end{figure}

Even though the data provide good coverage of the phase angle range [0$^\circ$,10$^\circ$], we did not attempt to constrain the phase function from the measured brightness, as we cannot disentangle the phase effect from intrinsic fading due to dust dissipation. Fig.~\ref{fig:reported_data} includes our measured apparent r-band magnitudes from ZTF out to the end of June 2023. Fig.~\ref{fig:ztf} shows our derived absolute magnitudes (again assuming $G$=0.20 $\pm$ 0.05) as a function of time out to late May, and linear fits for the r and g bands. 

\subsection{Archival 2024 ZTF images}
\label{sec:ZTF_2024}
           We also analysed ZTF images obtained at phase angles $<\!\!10^\circ$ between 03 August 2024 and 06 September 2024 (15 images in the g and r bands, respectively, and 7 in the i band, downloaded in snippets of 120\arcsec $\times$ 120\arcsec). We note that JN16 is barely visible in the individual images and we stacked the images to increase the signal-to-noise ratio (S/N). To account for the changing phase angle and heliocentric and geocentric distances, we subtracted a median background from each image and multiplied the result with $(r_{\rm h} \Delta)^2/\Phi(G,\alpha)$, using both $G$=0.15 and $G$=0.25. As the PA of the projected negative velocity vector changed by only 2.2$^\circ$, we did not rotate the images prior to stacking. We averaged all images obtained in the same filter in the co-moving frame of JN16 and rejected the 3 brightest values at each position to remove background stars. The resulting stacked images are displayed in Fig.~\ref{fig:2024_ZTF_tail} and show the tail, but not the nucleus, in all three filters. For calibration, we used the band-wise mean of all zeropoints from the image headers. 

Similar processing of data from the subsequent opposition (16 December 2025 -- 26 January 2026, 10 images in the g and r band, respectively) did not yield any visible trace of the tail, and the images are strongly affected by stellar contamination. We did not include these data into the subsequent analysis.

\subsection{Near-perihelion images from VLT 2024}
We were granted Director's Discretionary Time to observe JN16 in December 2024 with ESO's VLT+FORS2 and Bessel R Special filter (image scale 0.252\arcsec/pixel) to search for signs of renewed activity, such as a tail or coma of freshly emitted dust, or a nonstellar appearance of the nucleus brightness profile. The data cover 1h of observing time (13 images) on 02 December, and 0.5h (6 images) each on 30 and 31 December. The latter two dates bracket the perihelion passage. The individual images show a fuzzy dust tail of about 1\arcmin\ length, while the nucleus is not clearly visible. A journal of our VLT/FORS2 observations is given in Table~\ref{tab:vlt_obs}.

After bias subtraction, flat-fielding and background subtraction (approximating the background by the image median), we combined all images obtained during the same night in the co-moving frame of JN16, resulting in three stacked images, one from each night. Since the nucleus was not clearly visible in the single exposures, we stacked the images blindly, deriving the required offsets from the JPL Horizons\footnote{https://ssd.jpl.nasa.gov/horizons/app.html\#/} ephemeris (based on their orbit solution for 26 October 2024). To obtain a clean image of the dust tail, we rejected stars during the stacking process, exploiting the non-sidereal motion of the asteroid.

On 02 December, the apparent motion of the asteroid was parallel to the tail and the tail crossed two stars. To exclude these stars from the stacked image without removing flux originating from the tail, we proceeded as follows. We rotated all exposures such that the projected heliocentric velocity vector of JN16 (described by $PA_{-v}$ in Table~\ref{tab:vlt_obs} and almost aligned with the tail) was parallel to the horizontal axis of the rotated images. In each rotated image, we masked the tail with a rectangle of 285$\times$40 pixels, and averaged all images in the rest frame of the stars. This left a blank rectangle of 100$\times$40 pixels, in which the tail had been present in all single images. We filled this rectangle with column-wise interpolations of the background, after having verified that no bright stars had been present in this region. We then subtracted this stellar composite from each individual exposure and averaged the resulting frames in the co-moving frame of the asteroid, clipping all pixels significantly brighter than the brightest pixels of JN16, to further minimise stellar contamination.

\begin{figure*}
  \resizebox{\hsize}{!}{\includegraphics{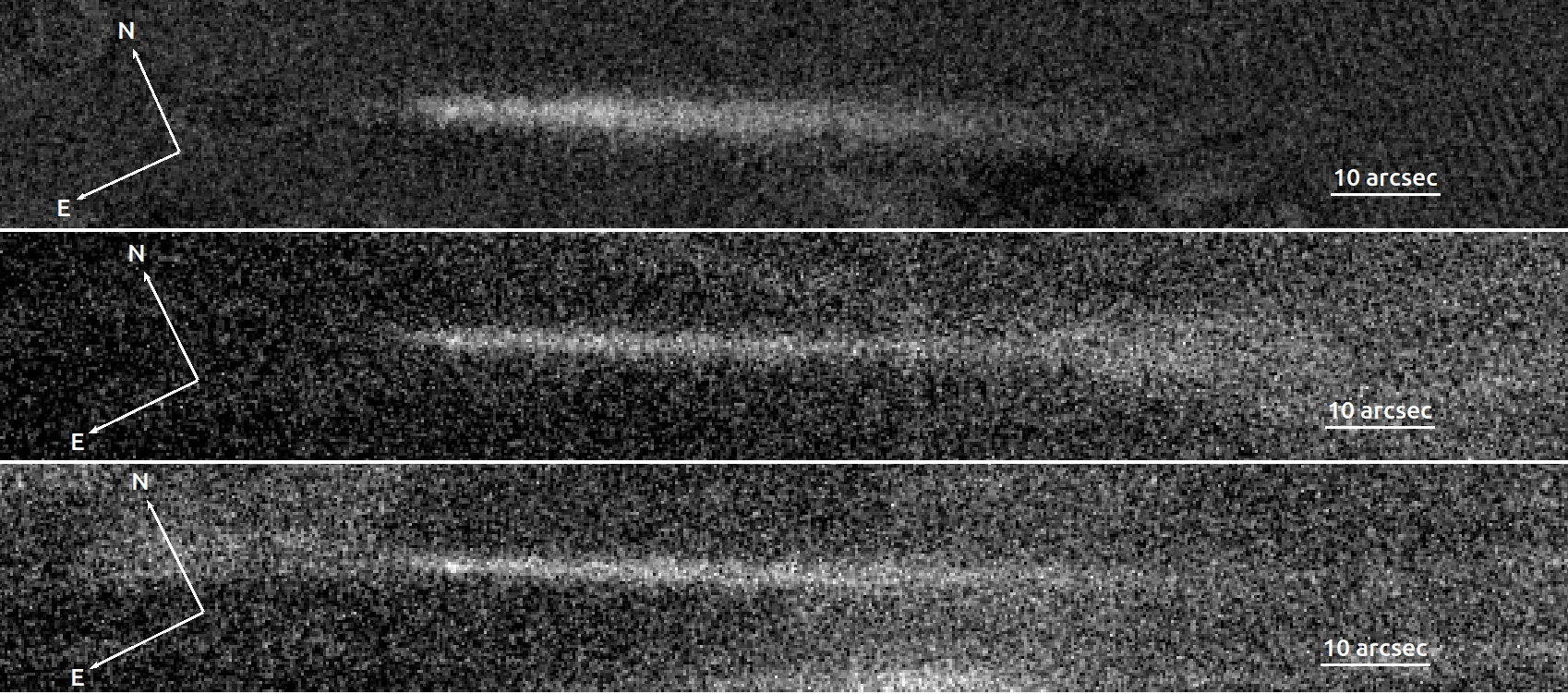}}
  \caption{Composite R-band images of the JN16 tail and nucleus from VLT/FORS2 from 02 (top), 30 (middle), and 31 (bottom) December 2024.}
  \label{fig:vlt_composites}
\end{figure*}

In the data from 30 and 31 December, the tail did not overlap with bright stars, such that we simply averaged all six frames from a given night in the co-moving frame of the asteroid while rejecting the three brightest pixels at each location. The three composite images are shown in Fig.~\ref{fig:vlt_composites}.

We photometrically calibrated each image using sunlike (AB system colours 0.2$<(g-r)<$0.6 and -0.08 $<(r-i)<$ 0.32) field stars from the Pan-STARRS DR2 catalogue\footnote{https://catalogs.mast.stsci.edu/panstarrs/}. To photometrically analyse the composite images, we used the weighted mean zeropoint of each night.

\section{Results}
\label{sec:results}

\subsection{Nucleus}
\label{sec:nucleus}
While the nucleus was not unambiguously detected in the individual frames, it can be discerned with sufficient confidence in the composites. In each of the three independent composites, the nucleus was located near the leading (in the sense of motion) end of the 3$\sigma$ position uncertainty ellipse of the JPL Horizons ephemeris (Figure~\ref{fig:nucleus}), rendering our identification of the nucleus robust. 

We measured the nucleus flux in a circular aperture of 3 pixel radius (green circles in Fig.~\ref{fig:nucleus}), and approximated the background by the mean flux in two circles of the same size positioned to the left and right of the nucleus (red circles in Fig.~\ref{fig:nucleus}). This background comprises both sky brightness and scattered light by the dust tail that extends to both sides of the nucleus. The measured apparent nucleus magnitudes are listed in Table~\ref{tab:vlt_nucleus} as well as the corresponding absolute magnitudes for the observing geometry shown in Table~\ref{tab:vlt_obs} and HG phase functions with $G$=0.15 ('C-type') and $G$=0.25 ('S-type'), respectively. For comparison, Table~\ref{tab:vlt_nucleus} also includes the serendipituous pre-discovery Subaru i-band measurement reported by \citet{CBET_JN16}.
We convert these absolute magnitudes to the product of the (unknown) geometric albedo, $p$, and the scattering cross-section, $C$, and to the albedo-dependent radius of a circle having the same cross-section according to
\begin{equation}
  r(p,H) = \frac{1 \mathrm{au}}{\sqrt{p}} 10^{-0.2(H - M_\odot)},
  \label{eq:radius}
\end{equation}
where $M_\odot$ is the absolute magnitude of the Sun in the appropriate filter band and we used PS1 values
$M^r_\odot$ = -26.93 and $M^i_\odot$ = -27.05 \citep{willmer2018}.

\begin{table*}[h]
 \caption{Nucleus properties before and after the dust emission event.}
\vspace{2mm}
\centering
\label{tab:vlt_nucleus}
\begin{tabular}{lllllllllll}
\hline\noalign{\smallskip}
Date [UT] & Band & $M_{\rm app}$ & $-\Delta M_{\rm app}$ & $+\Delta M_{\rm app}$ & $H^C$ & $H^S$ & $pC^C$ [km$^2$] & $pC^S$ [km$^2$] &  $r_n^C$[m] & $r_n^S$[m] \\
\hline\hline\noalign{\smallskip}
18 Mar 2018 & i & 24.7  & -0.1  & 0.1  & 20.52 & 20.54 & 0.0066$\pm$0.0006 & 0.0065$\pm$0.0006 & 187$\pm$9  & 91$\pm$4\\
02 Dec 2024 & r & 26.08 & -0.21 & 0.26 & 21.37 & 21.51 & 0.0034$\pm$0.0007 & 0.0030$\pm$0.0006 & 134$\pm$15 & 61$\pm$7\\
30 Dec 2024 & r & 25.86 & -0.20 & 0.25 & 20.93 & 21.06 & 0.0050$\pm$0.0010 & 0.0045$\pm$0.0009 & 163$\pm$17 & 75$\pm$8\\
31 Dec 2024 & r & 25.77 & -0.20 & 0.25 & 20.84 & 20.97 & 0.0055$\pm$0.0011 & 0.0049$\pm$0.0010 & 171$\pm$19 & 79$\pm$8\\
\noalign{\smallskip}\hline
\end{tabular}
\tablefoot{$M_{\rm app}$ is the apparent magnitude of the nucleus. The negative and positive error bars, $-/+\Delta M_{\rm app}$, correspond to the 1$\sigma$ statistical noise of the background. The 2018 i-band measurement was made with the Subaru telescope and reported in \citep{CBET_JN16}, while the three 2024 measurements derive from our VLT/FORS2 images. $H^{C/S}$ is the absolute magnitude in the concerned filter band assuming a C-type ($G$=0.15) and S-type ($G$=0.25) phase function, respectively. The quantity $pC$ represents the corresponding products of geometric albedo, $p$, and physical cross-section, $C$. The quantities $r_n^C$ and $r_n^S$ are the radii of a cross-section equivalent sphere assuming geometric albedos of $p$=0.06 for a C-type and $p$=0.25 for an S-type \citep{nugent-mainzer2016}. The uncertainty given for the radius derives from the statistical measurement uncertainty of the brightness, while the contribution of the unknown phase function is reflected by considering the 'S' and 'C' cases.}
\end{table*}

The nucleus radii derived from 2018 and two of the 2024 epochs are consistent within their error bars, while the value derived from 02 Decemeber 2024 is smaller, which can be due to rotational variability or imperfect subtraction of backgroud and/or of dust (Fig.~\ref{fig:reported_data}). The shape and rotation properties of the nucleus are unknown, and our data are insufficient to constrain them.
\citet{bottke-durda2005a} define a catastrophic disruption as an event where the largest fragment has $<$50\% of the original volume. Based on the pre-discovery size from Subaru, the $pC$ product of such a largest fragment would have to be $<0.0042$\,km$^2$, which is within the error bars of all three data points. Hence we cannot discriminate between a largely intact nucleus and the remnant of a catastrophic disruption, even less so as the intrinsic $r-i$ colour of JN16's surface and its shape are unknown.

However, we do not see any indication of renewed activity at perihelion, in which case the 2024 nucleus should have been surrounded by a bright (even if unresolved) coma of fresh dust. The absolute magnitudes derived from the central aperture of the 2023 ZTF images (Figure~\ref{fig:ztf}) are 5-6\,mag brighter, indicating that light scattering then was dominated by dust (see also Figure~\ref{fig:reported_data}). The surface escape speed, $v_{\rm esc} = \sqrt{4 \pi G \rho /3D} \, r_{\rm n}$, characterising a sphere of assumed density $\rho$ = 2000\,kg/m$^3$ and radius 61m$<r_{\rm n}<$187m, is 0.05\,m/s$<v_{\rm esc}<$0.14\,m/s.

\subsection{Dust cross-section}
\label{sec:dust_cross_sec}

We estimated the combined scattering cross-section of the nucleus and dust emitted in the time interval 27-29 April 2023 from the ZTF measurements of 29 April. These data were obtained closest in time to the supposed event (implying a minimum amount of loss from the employed aperture, see also Sect.~\ref{sec:dust_size_speed} below) and at the lowest available phase angle (implying a minimum uncertainty from the unknown phase function). We averaged the two r-band measurements to obtain an apparent r-band magnitude of $M_{\rm app}^r$ = 19.27$^{+0.16}_{-0.13}$. Table~\ref{tab:dust_mag} lists the apparent and absolute magnitudes (derived using Equation~\ref{eq:H}) of the dust and the corresponding products of geometric albedo and dust cross-section. We used a PS1 solar g-magnitude $M^g_\odot$ = -26.54 \citep{willmer2018}.

\begin{table*}[h]
\caption{Measured dust magnitudes and derived cross-sections.} 
\vspace{2mm}
\centering
\label{tab:dust_mag}
\begin{tabular}{lllllllllll}
\hline\noalign{\smallskip}
Date [UT] & Band & $M_{\rm app, d}$ & $-\Delta M_{\rm app, d}$ & $+\Delta M_{\rm app, d}$ & $H^C_{\rm d}$ & $H^S_{\rm d}$ & $pC^C$ & $pC^S$ & $-\Delta pC$ & $+\Delta pC$\\
\hline\hline\noalign{\smallskip}
29 Apr 2023 & g & 19.65 & -0.12 & 0.13 & 15.62 & 15.64 & 0.96 & 0.94 & -0.11 & 0.11\\
29 Apr 2023 & r & 19.27 & -0.13 & 0.16 & 15.24 & 15.26 & 0.95 & 0.94 & -0.13 & 0.12\\
03 Aug 2024 -- Sep-06 & g & --    & -0.07 & 0.07 & 16.24 & 16.24   & 0.54 & 0.54   & -0.04 & 0.03\\
03 Aug 2024 -- Sep-04 & r & --    & -0.04 & 0.04 & 15.51 & 15.51   & 0.74 & 0.74   & -0.03 & 0.03\\
03 Aug 2024 -- Sep-01 & i & --    & -0.04 & 0.04 & 15.17 & 15.17   & 0.90 & 0.90   & -0.03 & 0.03\\
02 Dec 2024 & r & 20.26 & -0.02 & 0.03 & 15.55 & 15.69 & 0.72 & 0.63 & -0.02 & 0.01 \\
30/31 Dec 2024 & r & 20.34 & -0.04 & 0.04 & 15.41 & 15.54 & 0.81 & 0.72 & -0.03 & 0.03\\
30/31 Dec 2024, small & r & 22.40 & -0.06 & 0.06 & 17.47 & 17.60 & 0.122 & 0.108 & -0.006 & 0.006\\
\noalign{\smallskip}\hline
\end{tabular}
  \tablefoot{Data in the first five lines are derived from ZTF images (stacked in the case of 2024). Data in the last three lines are derived from stacked VLT images, with the last two lines based on a combined stack of images from the two nights of 30 Dec and 31 Dec, and `small' refers to a rectangle extending to 5\arcsec\ on either side of the nucleus. $M_{\rm app, d}$ denotes the apparent magnitude of all light inside the considered aperture (see text for details), and its negative and positive error bars corresponding to the 1$\sigma$ statistical noise of the background are listed as $-/+\Delta M_{\rm app, d}$. Here, $H^{C/S}_{\rm d}$ is the absolute magnitude in the concerned filter band assuming a C-type ($G$=0.15) and S-type ($G$=0.25) phase function, respectively. The quantity $pC$ represents the product of geometric albedo, $p$, and physical cross-section, $C$, measured in km$^2$.}
\end{table*}

In the 2024 ZTF stacked images, we measured the integrated magnitude of the tail in a box of 80\arcsec $\times$ 9\arcsec, which covers the visible tail. We estimate the background as the average of the median values in boxes of the same size to either side of the tail (Fig.~\ref{fig:2024_ZTF_tail}). The measured absolute magnitudes are listed in Table~\ref{tab:dust_mag}, and the r-band value converted to an apparent magnitude for 20 August 2024 is shown in Fig.~\ref{fig:reported_data}. Within the given accuracy, C- and S-type $G$ parameters yield the same absolute magnitudes. We visually estimate the tail PA to be $(248 \pm 2)^\circ$ and add also this information to Fig.~\ref{fig:reported_data}. 

To constrain the dust cross-section in the stacked VLT images, we measured the flux in  70\arcsec$\times$6\arcsec\ rectangular apertures on the tail and derive the background estimate from rectangular apertures above and below the tail (Fig.~\ref{fig:tail}). The aperture size is chosen such that it emcompasses the visually detectable dust tail. The presence of additional dust outside this aperture (in particular to the west) is likely, but its surface brightness would be significantly below the noise level. The measured apparent and absolute magnitudes and corresponding $pC$ products are listed in Table~\ref{tab:dust_mag} and shown in Fig.~\ref{fig:reported_data}. At most one-third of the original dust cross-section has dissipated to locations outside our employed 2024 aperture. The contribution of the nucleus to the total scattering cross-section is of the order of 0.5\%.

\subsection{Tail orientation and width}
\label{sec:tail}
To better understand the ages, sizes, and velocities of the dust, we measured the orientation and width of the tail in the VLT images. For the first-order analysis, we compared the results to a calculation of the positions of dust particles ejected at zero relative speed as a function of their size and emission time.

Generally, the motion of dust particles, once they have left the source body's sphere of gravitational influence (Hill sphere), depends on their velocity upon gravitationally decoupling from the source and on the relative strength of solar radiation pressure and gravity, $\beta = C_\beta Q_{\rm pr}/(\rho s)$, where $\rho$ and $s$ are the bulk density and radius of the particle, respectively, $C_\beta = 5.77 \times 10^{-4}$, and $Q_{\rm pr}$ is the dimensionless radiation pressure coefficient \citep{burns-lamy1979}, assumed here to equal unity. For zero emission speed, the 3d-positions of the particles at any given time are uniquely related to the time of emission and their $\beta$ parameter. Following \cite{finson-probstein1968a} these positions can be visualised by a grid of synchrones (lines of constant emission time) and syndynes (lines of constant $\beta$). The assumption of zero speed is valid only if the influence of the initial speed, $v_{\rm ej}$, on the particle's orbital period is much smaller than that of radiation pressure.
In Section~\ref{sec:dust_size_speed} we evaluate if this criterion is fulfilled by the dust seen in our ESO data. 

\begin{figure}
  \resizebox{\hsize}{!}{\includegraphics{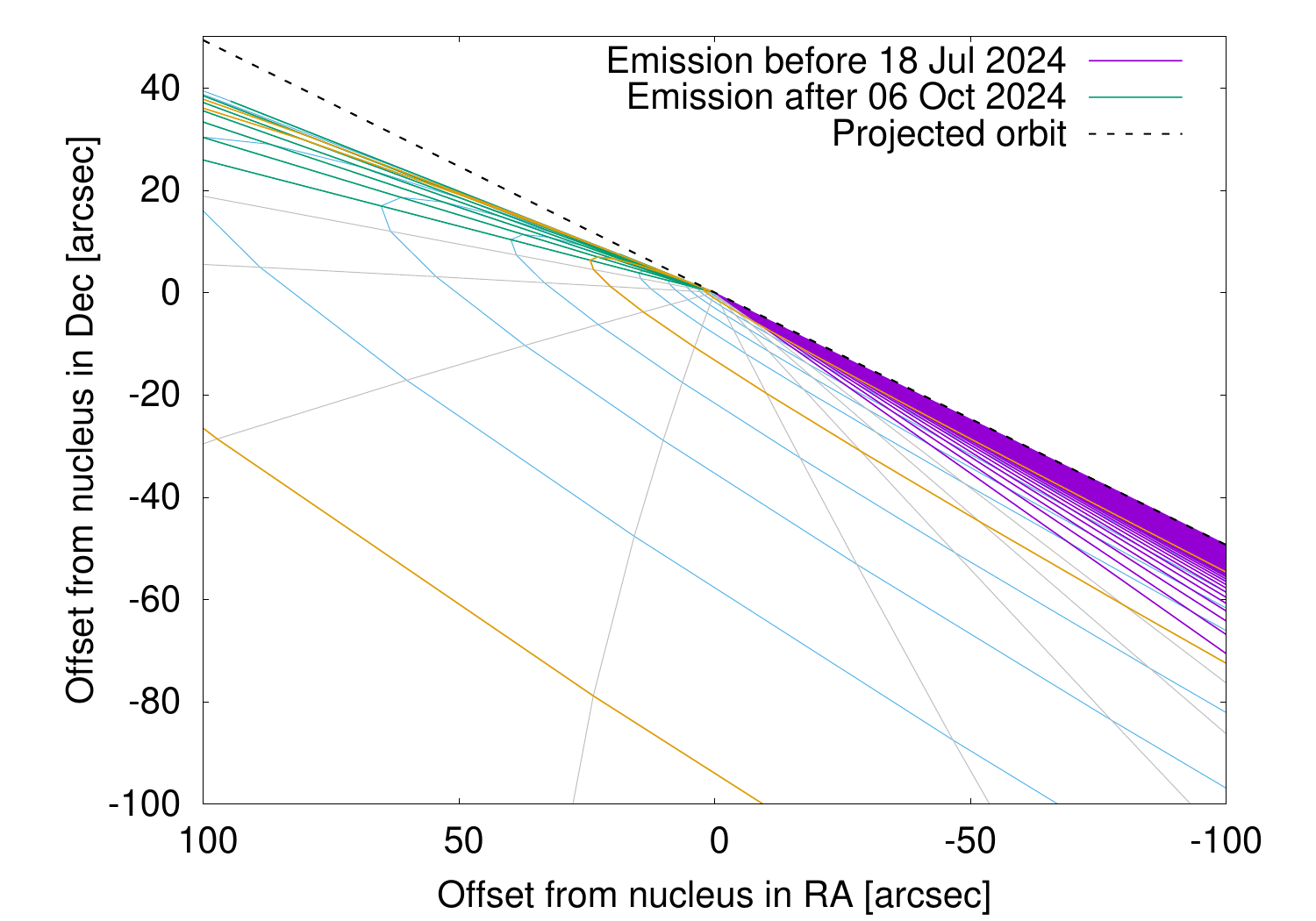}}
  \caption{Lines of constant ejection time (synchrones, violet, grey, and green) and constant radiation pressure parameter $\beta$ (syndynes, blue and orange) for an Earth-based observer on 30 December 2024. Orange lines correspond roughly to $\beta$ values of $10^{-3}, 10^{-2}, 10^{-1}$ as  nucleus distance increases. The dashed black line shows the PA of the projected heliocentric velocity vector of JN16, corresponding to its projected orbit.}
  \label{fig:synsyn}
\end{figure}

In Fig.~\ref{fig:synsyn}, we show a synchrone-syndyne grid for the circumstances of our second VLT epoch. Any dust ejected in late 2024 should be located east of the nucleus, while the dust we detect west of it would have been emitted before mid-2024. This dust should have $\beta<10^{-3}$. The data do show a short tail east of the nucleus, but we attribute this rather to non-zero ejection velocity (Section~\ref{sec:dust_size_speed}) than to freshly emitted dust, given the lack of brightening in the nucleus region (Section~\ref{sec:nucleus}).

\begin{figure}
  \resizebox{\hsize}{!}{\includegraphics{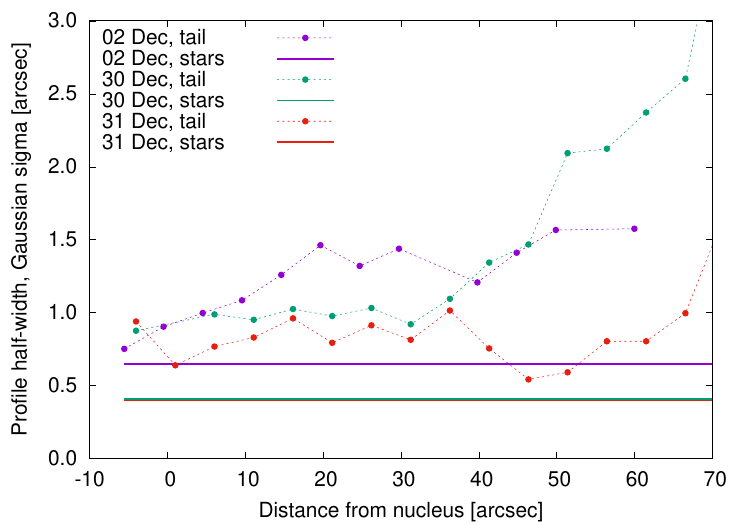}}
  \caption{Half-width (1$\sigma$) of Gaussian functions fitted to profiles measured perpendicular to the tail (averaging over 20px or 5\arcsec\ along the tail direction) and star trails (caused by the non-sidereal tracking of the telescope). The tail profiles were measured at various nucleus distances, while the width of stellar profiles is assumed independent of sky location. The stellar values represent the average of 5-7 star trails in each of the nightly sidereal composites.}
  \label{fig:width}
\end{figure}

To characterise the brightness distribution in the tail, we split the rotated images shown in Fig.~\ref{fig:vlt_composites} in vertical columns of 20px (5\arcsec) width and therein averaged each horizontal line of pixels, resulting in brightness profiles roughly perpendicular to the tail and spaced $\sim$5\arcsec\ apart. We fit each profile with a Gaussian function with the coordinate of the peak, the background level, and the width, $\sigma$, as free parameters. We also measured vertical profiles of star trails and fitted them with Gaussian functions to assess the contribution of the seeing to broadening the tail profiles. Fig.~\ref{fig:width} shows the resulting widths of the tail and star trails for all three VLT epochs.
The width of star trails indicates that the seeing was worse on 02 December than at the two later epochs. But at all three epochs, the dust tail is significantly broader (by about 0.5\arcsec\ or 2\,px in half-width) than the seeing profile, indicating that the physical extent of the projected tail is resolved in these images and covers about 1\arcsec. The deviation of the three tail profiles from each other beyond 40\arcsec\ indicates a strong influence of background inhomogeneities here. The non-zero width of the tail excludes that all dust was ejected both at a single point in time and with zero speed, in which case the tail should correspond to a single synchrone broadened only by seeing. We discuss the non-zero-speed interpretation in Section~\ref{sec:dust_size_speed}.

\begin{figure}
  \resizebox{\hsize}{!}{\includegraphics{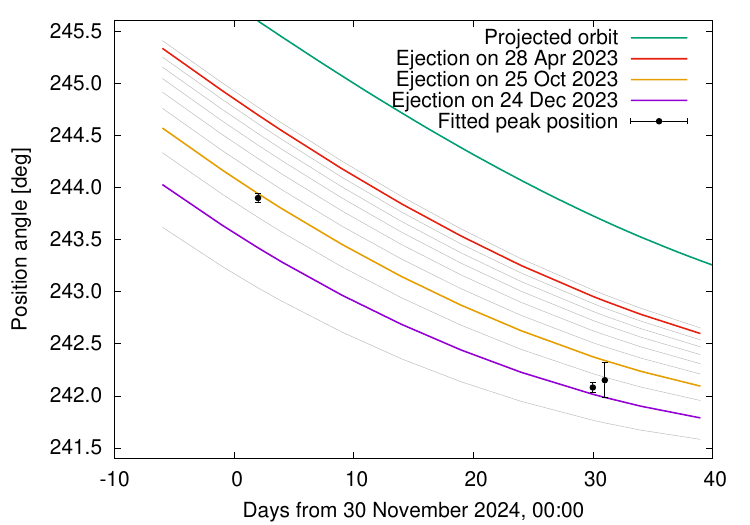}}
  \caption{PA as a function of time for a range of calculated synchrones (solid lines) and the tail observed with ESO/VLT. The emission time interval between neighbouring synchrones is 30\,days. Emission dates for the coloured synchrones are listed in the legend. The black symbols show the measured PAs as described in the text. The green line shows the PA of the projected negative heliocentric velocity vector of JN16, corresponding to its projected orbit.}
  \label{fig:pa}
\end{figure}

For zero speeds, the broadening of the tail would have to be attributed to an extended range of emission times, corresponding to a range in synchrones. At the relevant nucleus distances, synchrones are straight lines pointing radially away from the nucleus. At any given epoch, each synchrone is characterised by a unique sky PA. If the resolved width of the tail were caused by  an extended period of dust emission, it should increase with nucleus distance, for which the evidence in Fig.~\ref{fig:width} is marginal.
Fig.~\ref{fig:pa} shows the PAs of various synchrones as a function of observing time as continuous lines. The synchrone corresponding to emission on 29 April 2023 is marked in red. 
To assess the PA of the tail, we fit a linear function to the peak positions yielded by the Gaussian profiles as a function of nucleus distance. The resulting PAs are indicated by black symbols in Fig.~\ref{fig:pa}. This measurement is not consistent with the synchrone corresponding to emission on 28 April 2023. The measurements from all three epochs rather match synchrones in late 2023 (orange and violet lines in Fig.~\ref{fig:pa}). To exclude a systematic measurement error, we displayed the Gaia DR2 positions of field stars onto the single images according to the world coordinate system (WCS) information in the image headers. We did not find a systematic deviation of the calculated star positions from the observed positions that would indicate an erroneous sky rotation angle of the WCS. Also, the consistency of all three measurements with a similar range in synchrones at all three epochs at changing absolute angle makes an instrumental effect unlikely. We conclude that either the dust seen with VLT in December 2024 was not the same dust that was emitted during the April 2023 event or that the synchrone approximation is not valid due to a relevant influence of the emission speed on the heliocentric particle motion. During late 2023, the sky location of JN16 was rather close to the Sun, such that likely no archival data exist to assess a potential renewed activity in this time frame. However, if the dust seen by VLT had been emitted in late 2023, the 2023 ZTF dust would have to largely have left the FOV of VLT to explain the descrease in scattering cross-section between the two epochs. While we cannot exclude this, we think that the non-zero speed interpretation discussed in the next paragraph is more straightforward.

\subsection{Dust colour, sizes, and speed}
\label{sec:dust_size_speed}
We used multiple approaches to estimate the colours, sizes, and ejection velocities of the dust. First, we evaluated the fading of the absolute magnitude visible in Fig.~\ref{fig:ztf}. In May 2023, the phase angle increased steadily with time and since we do not know the dust phase function, we cannot tell if the fading is due to a potential overestimation of the value of the $G$ parameter or to the dissipation of dust out of the aperture. The linear fits suggest a slight change in colour (reddening, i.e. stronger fading in the g band), but a time-independent colour is also consistent with the error bars. If real, the reddening can either be attributed to a wavelength-dependent phase function or to an intrinsic change of dust particle properties (e.g. of their average size). Dissipation out of the aperture would primarily affect small particles that tend to have a more bluish colour, which would be qualitatively consistent with the observed reddening of the dust. 

The 2024 stacked ZTF images yield dust tail colours of $(g-r)$=(0.7 $\pm$ 0.1), $(r-i)$=(0.3 $\pm$ 0.1), and $(g-i)$=(1.1 $\pm$ 0.1), which is considerably redder than the 2023 measurement. Even though due to the low data quality and aggressive stacking procedure the true uncertainty of the measurement is likely larger than suggested by the formal error bars, the trend of reddening between 2023 and 2024 is rather clear. This may corroborate the hypothesis of reddening by loss of small dust.

\citet{colazo-alvarezcandal2022} and \citet{alvarezcandal2024} found that C-complex asteroids have $(g-i)<$0.7, while S- and X-complex objects have $(g-i)>$0.7. We conclude that our measured 2023 dust colours of 0.4$<(g-r)<$0.5 and $(r-i)\sim$0 are consistent with C-complex material, while the 2024 colours indicate S-type material. Therefore, in the following, we report all physical sizes both assuming C-type ($G$=0.15, $p$=0.06) and S-type ($G$=0.25, $p$=0.25) properties. The inconsistent attribution to spectral types between 2023 and 2024 may be an additional indicator that the decisive factor for dust colour in JN16 could be the size distribution rather than the material properties. 

Assuming that the (0.010 $\pm$ 0.003)\,mag/day fading of JN16 during May 2023 (Fig.~\ref{fig:ztf}) was due to the loss of dust scattering cross-section from the measurement aperture and using Eq.~\ref{eq:radius}, we find that during the interval $\Delta t_{\rm ZTF}$ =25 days  covered by ZTF data, about (20 $\pm$ 5)\% of the initial scattering cross-section have been lost from the central aperture. On 24 May 2023, the aperture radius of $\delta$=3\arcsec\ corresponded to about 4500\,km at the distance of the asteroid. Hence dust comprising (80 $\pm$ 5)\% of the scattering cross-section moved by $<$4500\,km in 25\,days. In the direction perpendicular to the orbital plane (where motion is hardly influenced by solar radiation pressure), this can be directly interpreted in terms of a speed: this dust had left the Hill sphere of JN16 at perpendicular speeds of $v_\perp <$ 2.1 m/s. 

\begin{figure*}[h]
  \includegraphics[width=0.5\textwidth]{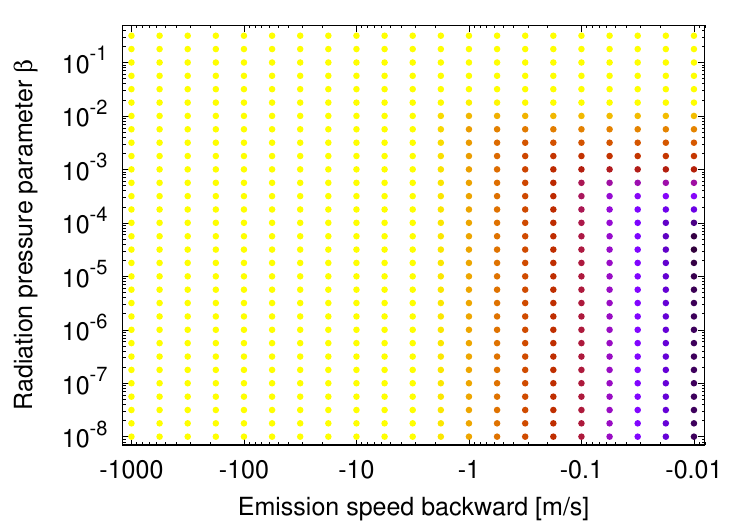}
    \includegraphics[width=0.5\textwidth]{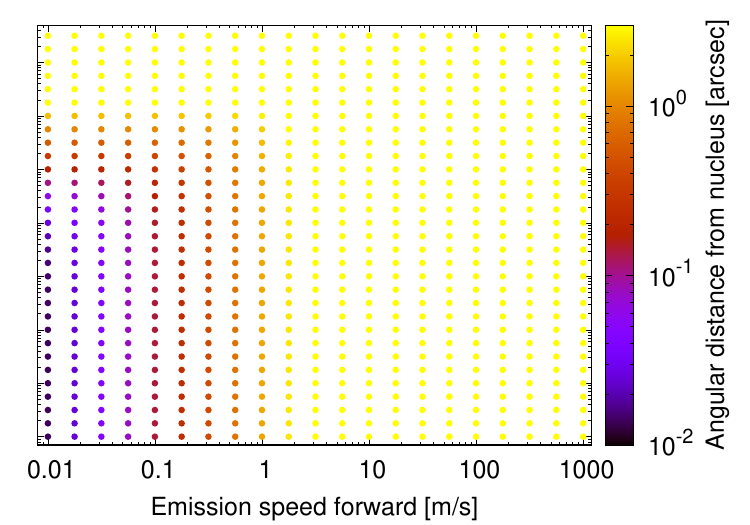}
    \caption{Angular separation from an asteroid (colour coded) as a function of emission speed parallel to the orbital motion of JN16 and radiation pressure parameter $\beta$ for particles emitted on 29 April 2023 and observed on 24 May 2023. Separations $>$3\arcsec\ (yellow points) correspond to locations outside the employed photometric aperture (Fig.~\ref{fig:ztf}). 
    }
  \label{fig:ztf_speed_beta}
\end{figure*}

Within the orbital plane, the dust motion relative to the nucleus is driven by the combined effect of solar radiation pressure and initial velocity. In Fig.~\ref{fig:ztf_speed_beta}, we calculated the angular nucleus distances on 24 May 2023 of dust particles emitted on 29 April 2023 in the direction parallel to the orbital velocity of JN16 with speeds, $v_{\parallel}$, ranging from $-$1000\,m/s (opposed to the direction of heliocentric motion) to $+$1000\,m/s and having $10^{-8}\leq\beta<0.4$.
Only particles with $|v_\parallel|<$2\,m/s and $\beta\leq 0.01$ (radius $s\geq$30$\mu$m assuming $\rho$=2000\,kg/m$^3$) will have remained inside the 3\arcsec\ aperture used to measure the brightness shown in Fig.~\ref{fig:ztf}. The small cross-section of (20$\pm$5)\% in particles with larger $\beta$ and/or speed is consistent with the absence of a visible dust tail in all 2023 ZTF images. 

\begin{figure*}[h]
  \includegraphics[width=0.5\textwidth]{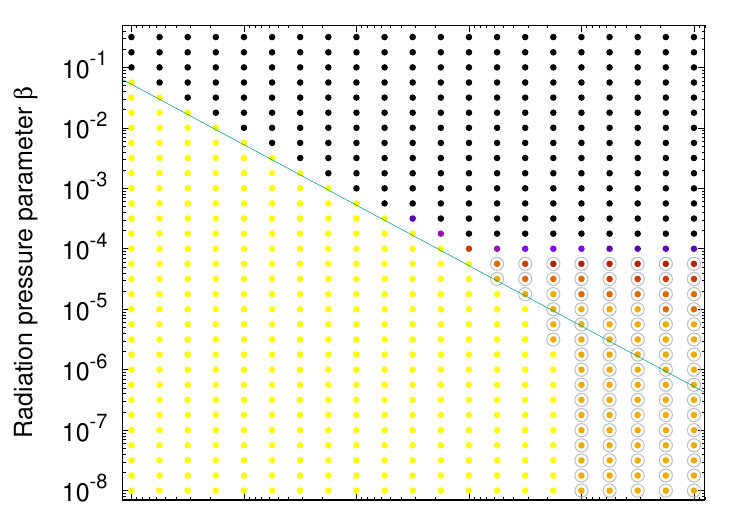}
  \includegraphics[width=0.5\textwidth]{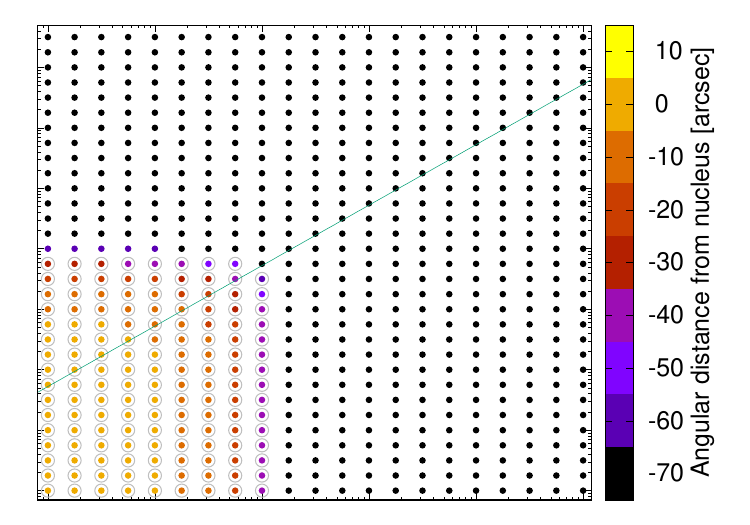}
   \includegraphics[width=0.5\textwidth]{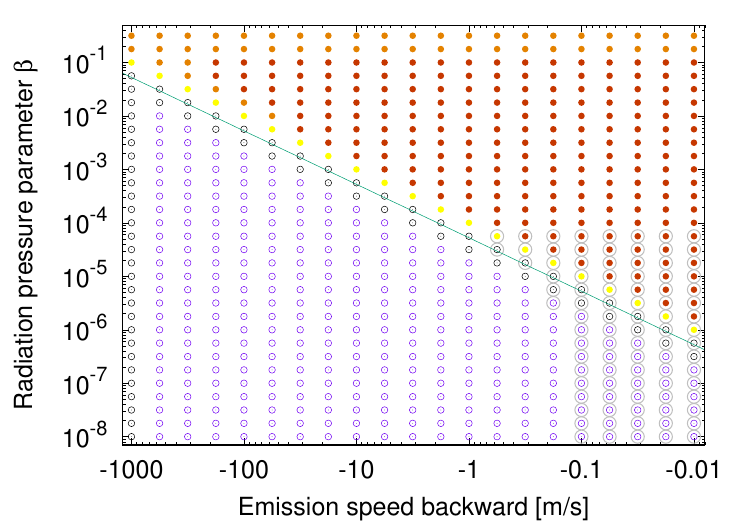}
  \includegraphics[width=0.5\textwidth]{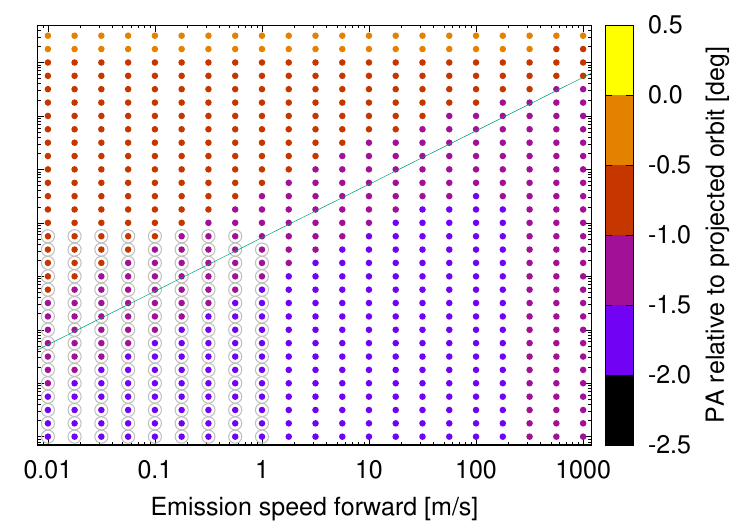}
     \caption{Angular separation from asteroid (top row) and PA (bottom row) as functions of emission speed parallel to the orbital motion of JN16 and radiation pressure parameter $\beta$ for particles emitted on 29 April 2023 and observed on 30 December 2024. The sign of the angular separation indicates whether a particle is projected to the east (positive values) or west of the nucleus. The PA is given relative to the PA of the projected orbit, 243.73$^\circ$. Coloured open symbols in the bottom row indicate locations east of the nucleus. Grey circles mark locations inside the observed dust tail. The green line represents the relation $\beta = |v_\parallel|$ / 19000\,m/s.}
  \label{fig:vlt_speed_beta}
\end{figure*}

We repeated this calculation for an observation on 30 December 2024 with the results shown in Fig.~\ref{fig:vlt_speed_beta}. Since we have a resolved image for this date, we also display the info related to whether a particle is east or west of the nucleus, along with the PA of a line connecting the projected positions of the particle and the nucleus. 
Grey circles in this plot mark combinations of $v_\parallel$ and $\beta$ that position a particle inside the visible tail, defined here as all combinations of $x$ and $y$ with $\sqrt{x^2+y^2} <$ 5\arcsec\ if east, or  $\sqrt{x^2+y^2} <$ 65\arcsec\ if west of the nucleus, $y < (0.5\arcsec + x \tan(270^\circ - (\overline{PA} + \Delta PA)))$, and $y > (-0.5\arcsec + x \tan(270^\circ - (\overline{PA} - \Delta PA)))$, where $\overline{PA}$=242.12$^\circ$ and $\Delta PA$=0.09$^\circ$ are the mean of the PAs measured on 30 and 31 December, and its error (Fig~\ref{fig:pa}), and we assume an intrinsic width of 1\arcsec\ (cf. Fig~\ref{fig:width}). 

In Fig.~\ref{fig:vlt_speed_beta}, we show that locations projected east of the nucleus are only reached by particles emitted to the direction opposed to the asteroid's orbital motion. Given that Earth was located south of JN16's orbital plane (angle $\epsilon<0$ in Table~\ref{tab:vlt_obs}), all dust located in the orbital plane and outside the orbit was projected to the south of the orbit, indicated by ($PA < PA_{\rm orb})$ in the west (closed symbols) and ($PA > PA_{\rm orb})$ in the east (not present in Fig.~\ref{fig:vlt_speed_beta}). Particles projected north of the projected orbit (all open and filled yellow symbols) are located inside the orbit of JN16. They are more tightly bound to the Sun then JN16 due to their reduced kinetic and total energy resulting from backwards emission. The eastern tail thus consists of particles ejected backwards mostly with $|v_\parallel| <$0.1\,m/s and $\beta < $10$^{-6}$, or up to 3$\times$10$^{-5}$ for specific speeds. Particles forming the western tail must have had $|v_\parallel| <$1\,m/s and $\beta < $10$^{-4}$.
In Sect.~\ref{sec:dust_cross_sec}, we explain how we estimated that no more than one-third of the original dust cross-section has dissipated to locations outside the visible tail, which we can now identify with particles having $\beta > 10^{-4}$ (radius $s\leq$3\,mm) and/or $|v_\parallel|>$1\,m/s.

These results show that the deviation of the measured tail PA from that of the synchrone of 28 April 2023 is aptly explained by a dominating influence of non-zero ejection speed over radiation pressure. To explore this further, we compared the difference in specific energy of a particle to that of the nucleus resulting from emission speed and solar radiation pressure. \cite{agarwal-jewitt2016} estimated that particle dynamics is dominated by radiation pressure if 
\begin{equation}
  v_{\rm e} \ll \beta \frac{G M_\odot}{r_{\rm h} v_{\rm n} \cos{\gamma}} = \beta \times 19\,000 \,\mathrm{m s^{-1}},
  \label{eq:v_beta}
  \end{equation}
where $r_{\rm h}$ and $v_{\rm n}$ are the heliocentric distance and speed of the nucleus at the time of dust emission, $G$ is the gravitational constant, $M_\odot$ is the mass of the Sun, and $\gamma$ is the angle between the heliocentric position vector and the ejection velocity, such that the ejection velocity component parallel to the heliocentric motion has the largest impact on a particle's trajectory. Fig.~\ref{fig:vlt_speed_beta} shows the relation $\beta = |v_\parallel|$ / (19000 m/s)  as green lines. For the angular distance (top row), the validity of Eq.~\ref{eq:v_beta} is well visible: above the line, distance depends on $\beta$, below the line on speed. By contrast, the sky PA (which is the quantity most often used to determine the emission date of a dust tail) seems to depend rather on the ratio of $\beta/v_\parallel$, and Eq.~\ref{eq:v_beta} being fulfilled is not sufficient for the synchrone approximation to be valid.

We generated plots similar to Fig.~\ref{fig:vlt_speed_beta}  for the emission directions perpendicular to the orbital motion of JN16 (Figs.~\ref{fig:vlt_speed_beta_perp} and \ref{fig:vlt_speed_beta_rad}). The resulting constraints for emission speed and radiation pressure parameter $\beta$ are similar to those arising from parallel emission. All visible particles must have $\beta<10^{-4}$ and emission speeds below 1\,m/s, where speeds exceeding 0.1\,m/s are allowed only in two of the six directions investigated (forwards and outwards, both increasing the total specific energy of the particle, making them less strongly bound to the Sun). In particular, only the fastest particles emitted forwards or inwards at about 1\,m/s or to the south at about 0.2\,m/s can explain the strong deviation from the projected orbit (-1.7$^\circ$) of the distant tail (closed violet symbols at same coordinates in both the upper and lower panels of Figs.~\ref{fig:vlt_speed_beta}, Figs.~\ref{fig:vlt_speed_beta_perp} and \ref{fig:vlt_speed_beta_rad}), while particles projected east of the nucleus must have been ejected backwards, inwards, or southwards at $v>$~$\beta$ 19\,000\,m/s.

\subsection{Excavated volume}
\label{sec:volume}
To translate the measured product of albedo and total cross-section into an estimate of the total volume of dust, the particle sizes would have to be known, and to translate this dust volume into the excavated volume, also knowledge of the macroporosity of surface material is required. Given the comparable influence of emission speed and solar radiation pressure, we cannot establish a unique relation between particle size and location in the tail, such that we neither know the particle size distribution nor the largest and smallest sizes.

We estimated from ZTF data that about 20\% of the total cross-section was in particles having a radius of $s<$30\,$\mu$ (unless their disappearance from the aperture was due to high speed). Comparing Figs.~\ref{fig:ztf_speed_beta} and \ref{fig:vlt_speed_beta}, we estimate that all particles seen in the VLT tail were also present in the ZTF aperture in May 2023, such that the particles lost between these two epochs (about 13\% of the cross-section) had speeds between 1\,m/s and 2\,m/s and/or sizes between 30\,$\mu$m and 3\,mm. From Figs~\ref{fig:vlt_speed_beta}, \ref{fig:vlt_speed_beta_perp}, and \ref{fig:vlt_speed_beta_rad} we infer that particles within 5\arcsec\ from the nucleus had $\beta<10^{-5}$ and $|v|<$0.1\,m/s, regardless of direction. This region contained about 12\% of the original cross-section (cf.~Table~\ref{tab:dust_mag}, last line). To derive a lower limit for the dust volume, we represent each of the discussed cross-section bins by the smallest possible particle size: 12\% of the cross-section was in particles with $s$=3\,cm ($\beta=10^{-5}$, near the nucleus in VLT), 55\% had $s$=3\,mm (the remaining VLT tail), 13\% had $s$=30\,$\mu$m (the dust lost between ZTF and VLT), and 20\% had $s$=3\,$\mu$m. The latter is an arbitrary size $<$30\,$\mu$m, but given that both the size and the contribution to the total cross-section are small, the uncertainty of this value does not much affect the resulting volume, which is between 27000\,m$^3$ ($p_{\rm S}$=0.25) and 111000\,m$^3$ ($p_{\rm C}$=0.06). This volume corresponds to 0.4\% (C) to 0.8\% (S) of the 2018 nucleus volume. It can be interpreted as a half-sphere of radius between $R_{\rm S}$=23\,m and $R_{\rm C}$=38\,m, or -- in the other extreme interpretation -- as a global layer of 25\,cm thickness on a spherical nucleus of 187\,m radius. The true excavated volume is likely larger than the dust volume because of non-zero macroporosity.

The above interpretation ascribes dust dissipation mainly to solar radiation pressure. However, if ascribed to ejection speed, the dust can also be much larger, with the size (and hence the volume) being unconstrained on the large end, as long as the nucleus cross-section does not change measurably. If all particles had $s$=1\,m, the debris volume would correspond to a global layer of 48\,m thickness, which is still just about consistent with the 2024 nucleus radii. Hence this marks our upper limit for the particle size and excavated volume.

\section{Discussion}
\label{sec:discussion}

\subsection{Cause of activity}
\label{sec:cause}
Given the lack of re-activation at perihelion and the near-aphelion location of the April 2023 event, we see no indications for the involvement of water ice sublimation in the dust emission from JN16. To assess whether a collision with another asteroid may have caused the 2023 dust emission, we compare JN16 to the dust emission history from three other main-belt asteroids that most likely or certainly underwent a collision and for which records of the subsequent dust evolution exist. These asteroids are (596) Scheila, 354P/LINEAR, and Dimorphos, the target of NASA's Double Asteroid Redirection Test \citep[DART,][]{daly-ernst2023}. The impact events onto Scheila and 354P were inferred from the very characteristic dust pattern in their tails that corresponded to radiation-pressure swept impact ejecta cones \citep{ishiguro-hanayama2011b, kim-ishiguro2017a}.

Scheila is considered of primitive spectral type \citep{bus-binzel2002b}, while  354P \citep{jewitt-ishiguro2013} and Dimorphos \citep[e.g. ][]{daly-ernst2023} are S-types. The pre-impact diameters of 354P \citep[sub-km-sized, $\gg$120\,m,][]{kim-ishiguro2017a, jewitt-weaver2010} and Dimorphos \citep[$\sim$150\,m,][]{daly-ernst2023} are well comparable to JN16. 
Scheila\footnote{https://ssd.jpl.nasa.gov/tools/sbdb\_lookup.html\#/?sstr=scheila} has a diameter of $>$100\,km  and an escape speed of 55\,m/s \citep{ishiguro-hanayama2011a}, which complicates a comparison with JN16. In addition, data on Scheila's dust scattering cross-section and dissipation are heterogeneous \citep[e.g.][]{larson2010, jewitt-weaver2011, ishiguro-hanayama2011a, neslusan-ivanova2016}. Telescope observations have witnessed particle sizes up to 100\,$\mu$m \citep{ishiguro-hanayama2011a}, while most large and slow ejecta would have fallen back to the surface without being observable to a telescope.

The asteroid 354P underwent a shattering collision, with the largest fragment being considerably smaller than the original body. However, no data exist on the latter, while the object was discovered only almost one year after the collision; thus, no information on the small dust is available, as it would have dissipated by the time of the discovery. 

Therefore, we chose to concentrate on the comparison with Dimorphos. The total cross-section of dust and debris ejected by the DART impact was 3.82\,km$^2$, of which 2.18\,km$^2$ left the immediate environment of the asteroid within 10\,min and the rest dissipated over the subsequent 18\,days \citep{li-hirabayashi2023, graykowski-lambert2023, kim-jewitt2023_dimorphos}. For JN16, we have no information on the brightness evolution on timescales of minutes or hours after the `event', but even after one month, the total cross-section in the central, unresolved aperture was about 5\,mag (a factor of 100) brighter than the nucleus, indicating a much higher fraction of large and slow ejecta. Consequently, Dimorphos built up a bright tail of small particles within days from the impact that was visible even by an 11cm-telescope with 1.3\,h of integration time \citep{graykowski-lambert2023}, while the first report of an extended tail at JN16 dates from 4 months after the event (Fig.~\ref{fig:reported_data}). On the other hand, the typical terminal speeds from Dimorphos \citep[ranging from near-zero to 1.5\,m/s,][]{kim-jewitt2023_dimorphos}, were comparable to those from JN16, and the overwhelming fraction of the mass ejected from Dimorphos was in metre-sized boulders \citep{jewitt-kim2023_dimorphos}, which we can neither confirm nor exclude for JN16. 

The other cause often invoked to explain short-duration, off-perihelion dust emission from asteroids is fast rotation, such that centrifugal acceleration overcomes gravity and cohesion in at least some regions on the surface \citep[e.g.][]{jewitt2012, jewitt_AIV}. The densest information supporting this hypothesis concerns five objects of three different types:

(1) Inner belt asteroids 6478 Gault and 311P emitted dust during more than one short-duration event on timescales of months. 
Gault has a radius of 1.4km \citep{devogele-ferrais2021} and a critically short rotation period of 2.5h \citep{luu-jewitt2021,purdum-lin2021, devogele-ferrais2021, carbognani-buzzoni2021}. Its rotation state is likely complex \citep{carbognani-buzzoni2021}.
Gault emitted dust in two episodes between October 2018 and January 2019 \citep{hui-kim2019}, and during at least three earlier epochs between 2013 and 2017 \citep{chandler-kueny2019}. The 2018/19 tails became visible to ZTF only about 1 month after the events \citep{ye-kelley2019}. Both these episodes showed the brightness in a central aperture rising for at least 10\,days per event \citep{ye-kelley2019}, suggesting that the emission process was not truly impulsive, unless the dust was initially optically thick or successively crumbling. The dust grain sizes inferred from radiation-pressure sweeping span the interval 20$\mu$m -- 3\,mm \citep{hui-kim2019} but the upper size limit is unknown due to the proximity that such particles would have kept to the nucleus. The dust speed has been estimated to (0.15$\pm$0.05)\,m/s \citep{jewitt_gault_2019,hui-kim2019}. The reported debris cross-section was 260\,km$^2$ for the first event, and 13\,km$^2$ for the second \citep{ye-kelley2019}. No tail was detected 20 months after the second event \citep[in August 2020,][]{purdum-lin2021}. The other asteroid in this group, 311P, has a radius of about 200\,m \citep{jewitt_P5-1}. Its rotation period is unknown, but it likely is a close binary system \citep{jewitt_P5-3}. In 2013, nine events of dust emission have been observed at irregular intervals over the course of nine months \citep{jewitt_P5-2}. The material from the first event (i.e. the tail with the highest synchrone age) was visible for 9\,months, while all subsequent tails faded below the detection limit after roughly 2\,months \citep{jewitt_P5-2}, indicating that the first event released larger particles than the subsequent ones. The largest particles were centimetre-sized ($\beta_{\rm min}<10^{-5}$), and no dust smaller than 10\,$\mu$m was detected ($\beta_{\rm max}=0.1$) \citep{jewitt_P5-2}. The out-of-plane terminal speed components are 0.06\,m/s, smaller than the escape speed $v_{\rm esc}$=0.27\,m/s \citep{jewitt_P5-2}. The reported total cross-section was 2\,km$^2$ \citep{jewitt_P5-1}. 

\begin{table*}
\caption{Key characteristics of some impacted and otherwise disrupted asteroids.}
\vspace{2mm}
\centering
\label{tab:observables}
\begin{tabular}{lrrrrrrrrrr}
\hline\noalign{\smallskip}
Object & $r_{\rm n}$[m] & $P$[h] & $\beta_{\rm min}$ & $\beta_{\rm max}$ & $\Delta m$ & $C$[km$^2$] & $f_{\rm month}$ & tail [months]& $v_{\rm max}$[m/s] \\
\hline\hline\noalign{\smallskip}
JN16 & 187 & -- &  $<$10$^{-4}$ & $<$0.01 & 5.5 & 15.8 & 0.8 & $>$20 & 1.0 \\

354P & $\gg$60\tablefootmark{a,b}  & 11.36\tablefootmark{b} & 10$^{-6}$\tablefootmark{c}& $>$7$\times$10$^{-5}$\tablefootmark{c} & -- & $>$13.9\tablefootmark{a} & -- & $>$43\tablefootmark{c} & $>$0.54\tablefootmark{c} \\ 

Scheila & $>$50$\times 10^3$ & 15.9\tablefootmark{d} & $<$10$^{-3}$\tablefootmark{h} & 0.3\tablefootmark{h} & $\sim$1\tablefootmark{e}& $\sim$10$^4$\tablefootmark{f}& $\sim$0\tablefootmark{f} & $>$3\tablefootmark{h} & $>$100\tablefootmark{g} \\

Dimorphos & $\sim$75\tablefootmark{i} & 11.9\tablefootmark{j} & $<$10$^{-6}$\tablefootmark{k} & 0.16\tablefootmark{k} & 5.8& 3.82\tablefootmark{l} & $\sim$0\tablefootmark{k,l} & $>$6\tablefootmark{m} & 1.5\tablefootmark{k} \\

311P & 200\tablefootmark{n} & -- & $<$10$^{-5}$\tablefootmark{o} & 0.1\tablefootmark{o} & $>$1.4\tablefootmark{o,p} & $>$2\tablefootmark{n} & -- & 9(2)\tablefootmark{o} & $>$0.06\tablefootmark{o} \\

Gault\tablefootmark{q} & 1400\tablefootmark{q} & 2.5\tablefootmark{q} & $<$2$\times 10^{-4}$\tablefootmark{r} & 0.04\tablefootmark{r} & 2.5\tablefootmark{s}& 260\tablefootmark{s} & -- & $<$20\tablefootmark{t} & 0.15\tablefootmark{r} \\

331P & 885\tablefootmark{u} & 3.24\tablefootmark{u} & $<$10$^{-9}$\tablefootmark{v} & 0.03\tablefootmark{w} & -- & $>$300\tablefootmark{w} & -- & $>$84\tablefootmark{v} & 0.1\tablefootmark{v} \\

62412 & 3900\tablefootmark{x} & 3.33\tablefootmark{x} & $<$10$^{-5}$\tablefootmark{y} & $>$10$^{-4}$\tablefootmark{y} & -- & -- & -- & 11 ... 20\tablefootmark{y} & -- \\

P/2013 R3 & -- & -- & $<$10$^{-9}$\tablefootmark{z} & 0.3\tablefootmark{$\alpha$} & -- & 30\tablefootmark{z} & 0.9\tablefootmark{z} & $>$5\tablefootmark{z} & 1\tablefootmark{$\alpha$}  \\
\noalign{\smallskip}\hline
\end{tabular}
\tablefoot{$r_{\rm n}$ and $P$ are the radius and rotation period of the nucleus or -- in case of 354P -- largest fragment. The $\beta$ values give the smallest and largest values in a continuous size distribution, inferred from dynamical modelling. $\Delta m$ is the magnitude increase by dust emission, corresponding to the dust and nucleus cross-section ratio. $C$ is the initial dust cross-section. $f_{\rm month}$ is the ratio of the brightness in a central aperture after one month over the initial dust brightness. Here, `tail' refers to the duration of visibility of the tail, while $v_{\rm max}$ is the speed of the fastest dust particles, inferred from modelling but using heterogeneous methods.
  \tablefoottext{a}{\!\!\citet{jewitt-weaver2010}}
  \tablefoottext{b}{\!\!\citet{kim-ishiguro2017b}}
  \tablefoottext{c}{\!\!\citet{kim-ishiguro2017a}}
  \tablefoottext{d}{\!\!\citet{warner2011}}
  \tablefoottext{e}{\!\!\citet{larson2010}}
  \tablefoottext{f}{\!\!\citet{jewitt-weaver2011}}
  \tablefoottext{g}{\!\!\citet{ishiguro-hanayama2011b}}
  \tablefoottext{h}{\!\!\citet{ishiguro-hanayama2011a}}
  \tablefoottext{i}{\!\!\citet{daly-ernst2023}}
\tablefoottext{j}{\!\!\citet{agrusa-gkolias2021}}
\tablefoottext{k}{\!\!\citet{kim-jewitt2023_dimorphos}}
\tablefoottext{l}{\!\!\citet{graykowski-lambert2023}}
\tablefoottext{m}{\!\!\citet{lister-constantinescu2024}}
\tablefoottext{n}{\!\!\citet{jewitt_P5-1}}
\tablefoottext{o}{\!\!\citet{jewitt_P5-2}}
\tablefoottext{p}{\!\!\citet{jewitt_P5-3}}
\tablefoottext{q}{\!\!\citet{devogele-ferrais2021}}
\tablefoottext{r}{\!\!\citet{hui-kim2019}}
\tablefoottext{s}{\!\!\citet{ye-kelley2019}}
\tablefoottext{t}{\!\!\citet{purdum-lin2021}}
\tablefoottext{u}{\!\!\citet{drahus-waniak2015}}
\tablefoottext{v}{\!\!\citet{jewitt-li2021}}
\tablefoottext{w}{\!\!\citet{stevenson-kramer2012}}
\tablefoottext{x}{\!\!\citet{sheppard-trujillo2015}}
\tablefoottext{y}{\!\!Mastropietro et al. (in prep.)}
\tablefoottext{z}{\!\!\citet{jewitt_R3-2}}
\tablefoottext{$\alpha$}{\!\!\citet{jewitt_R3}}
}
\end{table*}
(2) Outer belt, low-eccentricity asteroids 331P and 62412 are fast rotators \citep[periods $\sim$3.3\,h,][]{drahus-waniak2015, sheppard-trujillo2015} and emitted debris that was visible for at least one year. Both were discovered only almost one year after the dust emission events, and no data obtained closer in time to the events are available. 331P (diameter 1.77\,km) was discovered in March 2012, and the main dust emission was synchrone-dated to July 2011 \citep{stevenson-kramer2012}. The dust cross-section in 2012 was $>$300\,km$^2$, and the minimum dust size then was 20\,$\mu$m \citep{stevenson-kramer2012}. In 2015, the debris cross-section had reduced to 27\,km$^2$, and the largest unresolved debris was larger than centimetre-sized \citep{jewitt-li2021}. The debris emission speeds were about 0.1\,m/s. In addition to unresolved debris, individual (point-source) fragments have been detected with radii up to 111\,m \citep{jewitt-li2021}. \citet{hui-jewitt2022} inferred from the fragments' dynamical evolution that the fragmentation process was a cascade. 
The debris trail of asteroid 62412 was discovered in 2014 \citep{sheppard-trujillo2015}. Synchrones indicate that the trail at this time was at least 10 months old, and particle radii were $>$1\,mm. No information is available on the total scattering cross-section, but the absence of a gap between the trail and nucleus indicates that the largest particles were at least centimetre-sized.

(3) Outer belt asteroid P/2013 R3 disintegrated into at least 13 individually detected fragments (each about 200\,m radius) and debris over a five-month timescale in what was likely a cascading process. The relative speeds between fragments ranged between 0.2--0.5\,m/s, while the characteristic dust speed was estimated to 1\,m/s. The total dust cross-section was initially 30\,km$^2$, declining to 20\,km$^2$ within 105\,days. Micron-sized dust was initially present, and there is no reliable constraint on the maximum size \citep{jewitt_R3, jewitt_R3-2}

In Table~\ref{tab:observables}, we summarise some of the key observables at JN16, the three impacted and the five otherwise disrupted asteroids to assess if this comparison can yield any indication on the nature of the JN16 event. We find that collisions, if observed early, are characterised by a high fraction of small, fast dissipating dust, such that the central nucleus aperture returns to the pre-impact value within a month, which is different from JN16. We therefore think it unlikely that the cause of activity in JN16 was a collision. The comparison to the five other disrupted asteroids is hampered by an overall lack of consistent data, but based on the knowledge summarised in Table~\ref{tab:observables} we do not find a close similarity to group 1 (311P and Gault) because of the fast dissipation of their tails, indicating small particles, nor to P/2013 R3, which disintegrated. The closest similarity is with the fast rotators 331P and 62412, although no information is available on the initial presence of small dust in these two.

We speculate that the absence of small dust may be an intrinsic feature of JN16's surface material. Being rather small with weak gravity, JN16 and most of the other asteroids in Table~\ref{tab:observables} would not retain fine-grained ejecta from micro-meteorite impacts, which is the main source of the fine regolith on larger airless bodies such as Earth's moon \citep[e.g.][]{szalay-poppe2018}. Also in situ data of the 900m diameter primitive asteroid Ryugu from the Hayabusa 2 / MASCOT lander indicate the absence of significant deposits of fine-grained material \citep{jaumann-schmitz2019, grott-knollenberg2019} or put very stringent upper limits on their abundance \citep{hamm-hamilton2023}. It seems conceivable that the fine-grained material ejected by impacted asteroids or asteroids potentially experiencing multiple landslides or collisions within a binary system (suggested for 311P by \citet{hainaut-boehnhardt2014}) might be a by-product of these rather violent processes, or violent processes would be needed to detach fine dust from larger boulders to which it may be sticking by van-der-Waals forces. This would imply that the process leading to dust ejection from JN16 was less violent and created less fine dust. In this interpretation, and out of the commonly assumed triggers of asteroid activity, rotational mass shedding would seem the most likely cause of dust emission from JN16. This hypothesis requires to be tested by measuring JN16's rotational light curve.
The hypothesis of a gentle ejection process may be counter-indicated by the rather abrupt time evolution of the brightening.

Alternatively, the different debris sizes from Dimorphos (S-Type) and JN16 (if C-type) may be related to their potentially different composition. \cite{flynn-durda2005, flynn-durda2009} find that impact-generated dust from hydrous meteorites or meteorite analog material (related to C-complex asteroids) tends to have larger grain sizes than dust from anhydrous meteorites that relate to S-complex asteroids. A prerequisite with respect to pursuing this interpretation further would be a measurement of JN16's surface colour; however, in this paper, we only constrained the colours of single dust grains in the coma or tail.

\subsection{Constraints on ice content}
Based on the lack of re-activation at perihelion, we attempted to estimate the minimum depth of ice below the pre-event surface of JN16. This estimate is complicated by our lack of knowledge on (a) the nature of the event (i.e. if it is regional or global) and (b) the size change of the nucleus. The faintest constraint on the depth-to-ice would be given in a scenario where a global layer of debris would have been removed from a nucleus that remained otherwise intact. In this case, the ejected dust volume corresponds to a global layer of 25\,cm thickness for zero macroporosity (cf. Section~\ref{sec:volume}). This is comparable to the thermophysical constraint on the depth-to-ice in a non-sublimating asteroid. \citet{capria-marchi2012} have found that sublimation would still take place if the ice were covered by a $\Delta d$ = (10--50)\,cm thick layer of dry material, as long as that layer can be penetrated by the seasonal heat wave. Hence the minimum depth-to-ice in the pre-event JN16 must have been 25\,cm $+ \Delta d$. For an impact scenario, the minimum excavated dust volume would correspond to a bowl-shaped crater of 38\,m depth (Section~\ref{sec:volume}), a considerably stronger constraint.
Our estimated debris volume is a lower limit, because we do not know the abundance of metre- to decametre-sized fragments unresolved in the VLT images. In the most extreme interpretation, the 190\,m pre-event asteroid was reduced to a largest fragment of the order of 120--180\,m, possibly elongated, and an unknown quantity of smaller debris. In this case, the minimum depth-to-ice might approach 70\,m. 

\subsection{Activity in middle-main belt asteroids}

\begin{table*}
\caption{Reported occurrence of dust emission from small middle-main belt asteroids.}
\vspace{2mm}
\centering
\label{tab:middlebelt}
\begin{tabular}{llllll}
\hline\noalign{\smallskip}
Object & Recurrent & t.a. at   & t.a. of (discovery) & MPC data arc & Symbol in Fig.~\ref{fig:aa_aq} \\
       & activity? & discovery & activity\tablefootmark{$\alpha$}         &   &  and References     \\
\hline\hline\noalign{\smallskip}
JN16   & no        & 216$^\circ$ & 214$^\circ$ & Mar-2018 -- Jun-2024 & black\tablefootmark{a,b} \\
259P   & yes       & 18$^\circ$ & 315$^\circ$ -- 60$^\circ$ & Feb-2011 -- Sep-2017 & greed filled\tablefootmark{c,d}\\
457P   & yes       & 22$^\circ$ & 0$^\circ$ -- 52$^\circ$ & Apr-2015 -- Jun-2024 & green filled\tablefootmark{e} \\
2010 LH15 & yes & 10$^\circ$ & 325$^\circ$ -- 28$^\circ$ & Oct-2001 -- May-2023 & green filled\tablefootmark{f,g}\\
P/2019 A4 & ? & 10$^\circ$ & 357$^\circ$ -- 2$^\circ$ & Dec-2018 -- Mar-2019 & green open\tablefootmark{h,i}\\
P/2025 W4 & ? & 18$^\circ$ & $<$8$^\circ$\tablefootmark{$\delta$} & Dec-2025 -- Jan-2026 & green open\tablefootmark{q}\\
P/2015 X6 & no & 329$^\circ$ & 323$^\circ$ -- $>$342$^\circ$ & Nov-2015 -- Mar-2016  & red filled\tablefootmark{j,k,l,m}\\
2015 FW412 & no & 321$^\circ$ & $<$267$^\circ$\tablefootmark{$\beta$} & Mar-2015 -- Aug-2025 & red filled\tablefootmark{n}\\
P/2016 G1 & ? & 265$^\circ$ & 253$^\circ$ -- 258$^\circ$ & Mar-2016 -- Nov-2016 & red open\tablefootmark{o}\\
P/2025 W3 & ? & 82$^\circ$ & ?\tablefootmark{$\gamma$} & Nov-2025 -- Feb-2026 & red open\tablefootmark{p}\\
\noalign{\smallskip}\hline
\end{tabular}
\tablefoot{
  \tablefoottext{$\alpha$}{Column 4 shows the true anomaly range of recurrent activity or, for unconfirmed recurrency, the true anomaly at which the activity occured that led to the discovery.}
  \tablefoottext{a}{\!\!\citet{CBET_JN16}}
  \tablefoottext{b}{\!\!\citet{deen_mpec_JN16_aug24}}
  \tablefoottext{c}{\!\!\citet{garradd-sostero2008}}
\tablefoottext{d}{\!\!\citet{hsieh-ishiguro2021}}
\tablefoottext{e}{\!\!\citet{kim-jewitt2022}}
\tablefoottext{f}{\!\!\citet{chandler-oldroyd2023_LH15}}
\tablefoottext{g}{\!\!\citet{liu-li2025}}
\tablefoottext{h}{\!\!\citet{CBET_A4}}
\tablefoottext{i}{\!\!\citet{moreno-licandro2021}}
\tablefoottext{j}{\!\!\citet{CBET_X6}}
\tablefoottext{k}{\!\!\citet{moreno_X6}}
\tablefoottext{l}{\!\!\citet{hsieh-micheli2023}}
\tablefoottext{m}{\!\!\citet{martino-tancredi2025}}
\tablefoottext{n}{\!\!\citet{chandler-trujillo2023_FW412}}
\tablefoottext{o}{\!\!\citet{moreno_G1}}
\tablefoottext{p}{\!\!\citet{CBET_2025W3}}
\tablefoottext{q}{\!\!\citet{CBET_2025W4}}
\tablefoottext{$\beta$}{Based on reported tail PAs and synchrone-syndyne analysis by the author, using https://www.comet-toolbox.com/FP.html\#}
\tablefoottext{$\gamma$}{No synchrone-syndyne analysis possible, because PAs are degenerate.}
}
\end{table*}

Table~\ref{tab:middlebelt} gives an overview of the reported dust activity of small middle-belt asteroids. Out of the ten objects known to the author, three showed confirmed recurrent activity at perihelion. For three others (including JN16), re-activation at perihelion is very unlikely: For P/2015 X6, this conclusion was reached by \citet{hsieh-micheli2023} after an extensive but unsuccessful campaign to recover the object. For 2015 FW412, the author infers the low probability of perihelion activation from the temporal consistency of PanSTARRS photometry reported to the MPC\footnote{https://www.minorplanetcenter.net/db\_search/show\_object?object\_id=821467}. For the remaining four asteroids, data are not sufficient to answer whether or not they show recurrent activity. Table~\ref{tab:middlebelt} also lists the time interval (arc) for which the MPC database contains entries. A short arc of only the discovery epoch implies that the object was not (p)recovered during other apparitions and may mean that it did not again become as bright as during discovery. This may be an indirect indication of non-reactivation for at least P/2019 A4 and P/2016 G1, while the discoveries of P/2025 W3 and P/2025 W4 are too recent for a meaningful interpretation of the data arc. As far as a statistical interpretation of such a small sample is meaningful, we conclude that a substantial fraction ($\gtrsim$ 30\%) of activated middle-belt asteroids do not subsequently take up sublimation-driven activity. An interesting case is P/457P, that showed recurrent dust emission near perihelion, but for the water production rate of which JWST spectra obtained near perihelion ($\nu$=9$^\circ$) yielded only a strong upper limit of $<2\times10^{24}$\,molecules/s \citep{noonan-hsieh2026}. This is about a factor ten below the typical production rate measured by JWST in three MBCs of the outer main belt \citep{kelley-hsieh2023, hsieh-noonan2025, hsieh-noonan2026}. \citet{noonan-hsieh2026} propose that the activity of 457P may be confined to a small region ($<$50\,m), or else that it had already ceased by the time of observation. Both \citet{kim-jewitt2022} and \citet{noonan-hsieh2026} found indications of the absence of particles smaller than millimetre-sized, which is in line with our findings for JN16.

\subsection{Limits of synchrone models for large debris}
To reconstruct the emission date of debris in narrow asteroid tails, the synchrone interpretation \citep{finson-probstein1968a} is often used, in particular when measurements of the tail PAs at multiple times are available \citep[e.g.][]{jewitt-weaver2010, stevenson-kramer2012, jewitt_P5-2, jewitt_gault_2019}. This approximation is strictly valid only for zero emission speeds. \cite{agarwal-jewitt2016} derived an expression estimating when the zero-speed assumption can be considered as a valid approximation (Eq.~\ref{eq:v_beta} above) from a consideration of specific energy differences between asteroid and dust particle. In active asteroids this conditions is more likely to be fulfilled than in classical comets due to typically smaller ejection speeds. But the validity of Eq.~\ref{eq:v_beta} can often not be assessed, when neither the dust speeds nor the date of emission are known and the synchrone method is the only means to constrain the date of emission. JN16 gives us the opportunity to assess the validity of the zero speed assumption because we have an independent constraint of the date of dust ejection from the ZTF archive. We found in Sect.~\ref{sec:tail} that the synchrone date inferred from the measured tail PAs differs from the ZTF-derived date by 6 -- 8 months, and showed in Sect.~\ref{sec:dust_size_speed} that the ZTF-derived date can be reconciled with the tail orientation in the VLT images only if Eq.~\ref{eq:v_beta} is largely not fullfilled (see all combinations of speed and $\beta$ below the green line in Fig.~\ref{fig:vlt_speed_beta}), i.e. that dust particles were sufficiently large for the $<$1\,m/s speeds to measurably influence their trajectories. This illustrates that synchrone-dating dust emission events should be used with care, especially in cases of long-lived dust tails that remain near the source object on timescales of years. We assume that the synchrone dates derived for 311P \citep{jewitt_P5-2} and Gault \citep{hui-kim2019} are not so much affected by this limitation, because the dust was small enough to dissipate on timescales of about two months. By contrast, the tails of 331P \citep{stevenson-kramer2012} and 62412 \citep{sheppard-trujillo2015} were discovered only about one year after the respective dust emission events (from synchrone dating). Hence the observed dust was clearly large and comparable in size to that from JN16. Since no data from closer to these events are known, we cannot assess if the zero-speed approximation was valid in these cases.

\section{Summary}
\label{sec:conclusion}
We analysed ZTF images of active asteroid P/2023 JN16 from August to September 2024, along with ESO/VLT/FORS2 images obtained around its perihelion passage in December 2024. We also re-analysed ZTF images of the same asteroid obtained during the first half of 2023. We measured the brightness of dust and nucleus and interpret these measurements through comparison to computer simulations of the dust motion under the influence of solar gravity and radiation pressure, and as a function of emission speed from the nucleus. Our key findings are summarised as follows.

\begin{itemize}
\item We confirm the finding by \citet{CBET_JN16} and \citet{deen_mpec_JN16_aug24} that middle main-belt asteroid P/2023 JN16 emitted dust between 27 and 29 April 2023, causing it to brighten by about five to six magnitudes.
\item During the month following the activation event, the dust colour was roughly neutral, consistent with C-complex material; whereas in 2024, it was redder and rather consistent with S-complex material. It is possible that the predominant loss of small particles from the dust ensemble contributed to the reddening trend.
\item The original nucleus (serendipituously observed by Subaru in 2018 and photometrically measured by \citet{deen_mpec_JN16_aug24}) had a cross-section equivalent radius of (187$\pm$9)\,m for a C-type albedo of 6\%. Post-event (2024) radii are smaller, but their large variability and measurement uncertainties make them compatible with both a catastrophic disruption scenario and a largely intact nucleus. The surface escape speed from a 187\,m radius sphere of assumed density 2000\,kg/m$^3$ is 0.14\,m/s.
\item According to our computer simulation of the dust motion, most of the debris cross-section was in particles having radii $s>$3\,mm, and compatible with up to metre sizes, and left the Hill sphere of JN16 with speeds of 1\,m/s and less. 
\item Our lower limit estimate of the excavated volume corresponds to a semi-spherical crater of 38\,m depth (assuming C-type albedo) or a global layer of 25\,cm, or anything in between.
  On the other extreme, assuming mainly metre-sized fragments, we cannot exclude a catastrophic disruption, with the post-event JN16 having $<$50\% of its original volume.
\item We did not find any indications of renewed (sublimation-driven) activity during the 2024 perihelion. We conclude that the 2023 event did not excavate ice. The minimum depth of ice on JN16 is 35-75\,cm, when taking into account that even after the dust removal, a sufficiently thick layer of dry material is required to prevent ice from sublimating.
\item From comparison to impacted asteroids Scheila and Dimorphos, we do not favour an impact as the dust-emitting event on JN16 due to the absence of small, fast-dissipating dust in JN16.
\item Synchrone-dating the dust emission event mismatches the date of sudden brightening by six to eight months due to the non-validity of the zero-speed approximation, indicating the limits of this method for the age determination of large-particle debris trails.
\end{itemize}
Further progress in understanding the circumstances of dust emission from JN16 would benefit from the following additional measurements. (1) The rotation period and lightcurve amplitude will help to assess the probability of rotational mass shedding and constrain the change in nucleus size by the 2023 event. (2) Spectral colours of the nucleus surface would establish the spectral type of JN16 and inform whether this may be an explanation for the lack of small particles compared to other impacted asteroids. (3) Continued monitoring of the debris field will provide additional constraints to the size of large debris and hence to the excavated volume.

\begin{acknowledgements}

We thank the anonymous referee for their comments that helped to improve the quality of this manuscript.
J.A. acknowledges funding by the  Volkswagen Foundation.
J.A. thanks the Max Planck Institute for Solar System Research in G\"ottingen for access to its computing facilities.
The results presented in this work are based on observations collected at the European Southern Observatory under ESO Director's Discretionary Time programme(s) 114.28HC.001.
  This research is in part based on data obtained by the Zwicky Transient Facility and provided by the NASA/IPAC Infrared Science Archive, which is funded by the National Aeronautics and Space Administration and operated by the California Institute of Technology.
This work has made use of data from the Pan-STARRS1 Survey (PS1) and the PS1 public science archive accessed through the Space Telescope Science Institute (STScI) / Barbara A. Mikulski Archive for Space Telescopes (MAST). 
The Pan-STARRS1 Surveys (PS1) and the PS1 public science archive have been made possible through contributions by the Institute for Astronomy, the University of Hawaii, the Pan-STARRS Project Office, the Max-Planck Society and its participating institutes, the Max Planck Institute for Astronomy, Heidelberg and the Max Planck Institute for Extraterrestrial Physics, Garching, The Johns Hopkins University, Durham University, the University of Edinburgh, the Queen's University Belfast, the Harvard-Smithsonian Center for Astrophysics, the Las Cumbres Observatory Global Telescope Network Incorporated, the National Central University of Taiwan, the Space Telescope Science Institute, the National Aeronautics and Space Administration under Grant No. NNX08AR22G issued through the Planetary Science Division of the NASA Science Mission Directorate, the National Science Foundation Grant No. AST–1238877, the University of Maryland, Eotvos Lorand University (ELTE), the Los Alamos National Laboratory, and the Gordon and Betty Moore Foundation.
This work has made use of astrometric results from the European Space Agency (ESA) space mission Gaia. Gaia data are being processed by the Gaia Data Processing and Analysis Consortium (DPAC). Funding for the DPAC is provided by national institutions, in particular the institutions participating in the Gaia MultiLateral Agreement (MLA). The Gaia mission website is https://www.cosmos.esa.int/gaia. The Gaia archive website is https://archives.esac.esa.int/gaia.
This work made use of the software IRAF, distributed by the Community Science and Data Center at NSF NOIRLab, which is managed by the Association of Universities for Research in Astronomy (AURA) under a cooperative agreement with the U.S. National Science Foundation.
This work has made use of the software SAOImageDS9, which is fully funded by the Chandra X-ray Science Center (CXC) and is licensed in part under the GNU General Public License, version 3.
This research has made use of NASA's Astrophysics Data System, of the JPL/Horizons ephemerides service, maintained by the JPL Solar System Dynamics group, and of data and/or services provided by the International Astronomical Union's Minor Planet Center. 
\end{acknowledgements}

\bibliographystyle{aa} 
\bibliography{refs} 

\begin{appendix}
\onecolumn
  
\section{Observational circumstances of the images analysed}

Table~\ref{tab:ztf} shows the parameters of and photometric results from ZTF images of JN16 used in this paper. Table~\ref{tab:vlt_obs} shows the circumstances and parameters of our DDT observations with ESO/VLT/FORS2 from December 2024.
\begin{table*}[h!]
  \caption{Parameters of the ZTF images covering JN16 between 27 April 2023 and 25 May 2023.} 
\vspace{2mm}
\centering
\label{tab:ztf}
\begin{tabular}{llrrrrrrrrrr}
\hline\noalign{\smallskip}
Date [UT] & Band & $r_h$ & $\Delta$ & $\alpha$ & $PA_{-\odot}$ & $PA_{-v}$ & $\epsilon$ & $\nu$ & $M_{\rm app}$ & -$\Delta M_{\rm app}$ & +$\Delta M_{\rm app}$ \\
\hline\hline\noalign{\smallskip}
 27 Apr 2023, 07:05:54 & r & 3.007 & 2.002 & 1.22 & 331.69 & 292.16 & 0.78 & 213.44 & $>$20.55 &&\\
 27 Apr 2023, 08:16:36 & g & 3.007 & 2.002 & 1.21 & 332.34 & 292.17 & 0.78 & 213.45 & $>$20.85 &&\\
 29 Apr 2023, 06:56:51 & r & 3.005 & 1.999 & 0.85 &  12.60 & 292.31 & 0.84 & 213.80 & 19.29 & -0.12 & 0.13 \\
 29 Apr 2023, 07:35:10 & r & 3.005 & 1.999 & 0.84 &  13.33 & 292.31 & 0.84 & 213.80 & 19.25 & -0.23 & 0.29 \\
 29 Apr 2023, 10:06:07 & g & 3.005 & 1.999 & 0.84 &  16.18 & 292.31 & 0.84 & 213.82 & 19.65 & -0.13 & 0.15 \\
 01 May 2023, 07:06:14 & g & 3.004 & 1.997 & 1.10 &  58.93 & 292.45 & 0.89 & 214.15 & 19.65 & -0.21 & 0.25 \\
 12 May 2023, 07:50:40 & r & 2.994 & 2.008 & 5.14 &  99.83 & 293.21 & 1.18 & 216.11 & 19.70 & -0.41 & 0.67 \\
 14 May 2023, 06:32:23 & r & 2.992 & 2.013 & 5.88 & 101.17 & 293.34 & 1.22 & 216.46 & 19.71 & -0.24 & 0.31 \\
 14 May 2023, 07:35:50 & g & 2.992 & 2.014 & 5.90 & 101.20 & 293.34 & 1.22 & 216.47 & 20.23 & -0.31 & 0.44 \\
 15 May 2023, 06:33:00 & i & 2.991 & 2.017 & 6.26 & 101.75 & 293.40 & 1.24 & 216.64 & 19.65 & -0.17 & 0.21 \\
 16 May 2023, 06:05:43 & g & 2.990 & 2.020 & 6.62 & 102.26 & 293.46 & 1.27 & 216.82 & 20.18 & -0.15 & 0.18 \\
 20 May 2023, 05:30:58 & r & 2.986 & 2.036 & 8.08 & 103.89 & 293.70 & 1.34 & 217.53 & 19.81 & -0.11 & 0.12 \\
 20 May 2023, 07:02:17 & i & 2.986 & 2.037 & 8.10 & 103.91 & 293.71 & 1.34 & 217.54 & 20.02 & -0.24 & 0.30 \\
 20 May 2023, 07:35:03 & g & 2.986 & 2.037 & 8.11 & 103.92 & 293.71 & 1.35 & 217.54 & 20.37 & -0.32 & 0.45 \\
 22 May 2023, 06:01:10 & g & 2.985 & 2.046 & 8.79 & 104.53 & 293.82 & 1.38 & 217.89 & 20.50 & -0.17 & 0.21 \\
 22 May 2023, 08:14:26 & r & 2.984 & 2.047 & 8.83 & 104.56 & 293.82 & 1.38 & 217.91 & 19.79 & -0.11 & 0.13 \\
 24 May 2023, 06:07:49 & r & 2.983 & 2.057 & 9.49 & 105.08 & 293.92 & 1.41 & 218.25 & 20.14 & -0.13 & 0.14 \\
 24 May 2023, 07:03:09 & g & 2.983 & 2.057 & 9.50 & 105.09 & 293.93 & 1.41 & 218.26 & 20.39 & -0.16 & 0.19 \\
 25 May 2023, 04:31:16 & i & 2.982 & 2.062 & 9.80 & 105.31 & 293.97 & 1.43 & 218.42 & 19.64 & -0.18 & 0.21 \\ 
\noalign{\smallskip}\hline
\end{tabular}
\tablefoot{All images had exposure times of 30\,s. $r_h$ and $\Delta$ are the heliocentric and geocentric distances in au, $\alpha$ is the phase angle, $PA_{-\odot}$ and $PA_{-v}$ are the PAs of the anti-solar direction and of the projected negative orbital velocity vector, $\epsilon$ is the angle between the line of sight and the heliocentric orbital plane of the asteroid, and $\nu$ is the true anomaly angle. All angles are measured in degrees. The last three columns list the apparent magnitude and its negative and positive error bars corresponding to the 1$\sigma$ statistical noise of the background.}
\end{table*}
\begin{table*}[h]
\caption{Parameters of the archival 2018 Subaru images and of our VLT/FORS2 images obtained from ESO programme 114.28HC.001.} 
\vspace{2mm}
\centering
\label{tab:vlt_obs}
\begin{tabular}{llrrrrrrrrrr}
\hline\noalign{\smallskip}
Date [UT] & $T_{\rm exp}$ & $N$ & $r_h$ & $\Delta$ & $\alpha$ & $PA_{-\odot}$ & $PA_{-v}$ & $\epsilon$ & $\nu$ \\
\hline\hline\noalign{\smallskip}
18 Mar 2017, 11:06:44 & 200 &  3 & 3.091 & 2.096 &  0.83 & 148.65 & 296.96 & -0.45 & 171.61\\
02 Dec 2024, 00:36:17 & 225 & 13 & 2.303 & 2.234 & 25.04 &  69.25 & 245.60 & -1.38 & 351.36\\
30 Dec 2024, 01:07:50 & 215 &  6 & 2.300 & 2.557 & 22.57 &  67.74 & 243.73 & -1.41 & 359.83 \\
31 Dec 2024, 00:55:56 & 215 &  6 & 2.300 & 2.568 & 22.44 &  67.69 & 243.68 & -1.41 & 0.13\\
\noalign{\smallskip}\hline
\end{tabular}
\tablefoot{The dates given are for the start of the first exposure in a given night. $T_{\rm exp}$ is the exposure time in seconds, $N$ is number of images, $r_h$ and $\Delta$ are the heliocentric and geocentric distances in au, $\alpha$ is the phase angle, $PA_{-\odot}$ and $PA_{-v}$ are the PAs of the anti-solar direction and of the projected negative orbital velocity vector, $\epsilon$ is the angle between the line of sight and the heliocentric orbital plane of the asteroid, and $\nu$ is the true anomaly angle. All angles are measured in degrees. All VLT images were obtained with the Bessel R Special filter, the Subaru images were taken in the i band.}
\end{table*}

\section{Apertures used for nucleus and dust photometry on composite images} 
\begin{figure}
  \resizebox{\hsize}{!}{\includegraphics{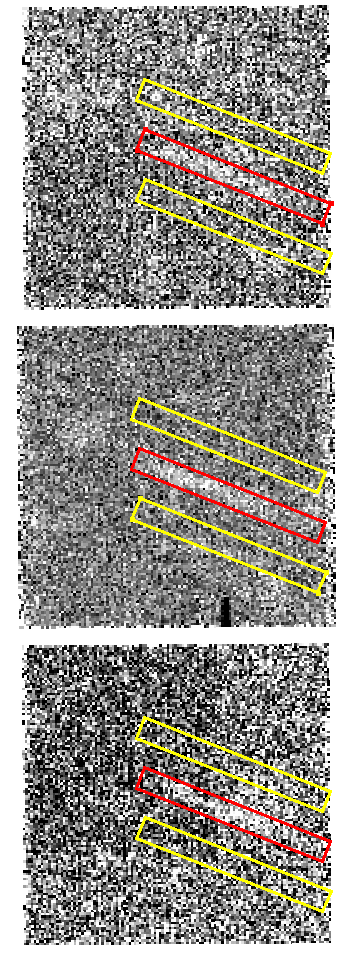}}
\caption{Stacked ZTF images obtained between 03 August 2024 and 06 September 2024 in the g band (top), r band (middle), and i band (bottom). Each image has a linear dimension of 125\arcsec, North is up and east is to the left. The red box was used to measure the integrated flux from the tail, while the median brightness in the yellow boxes served to estimate the background.}
  \label{fig:2024_ZTF_tail}
\end{figure}

\begin{figure}
  \resizebox{\hsize}{!}{\includegraphics{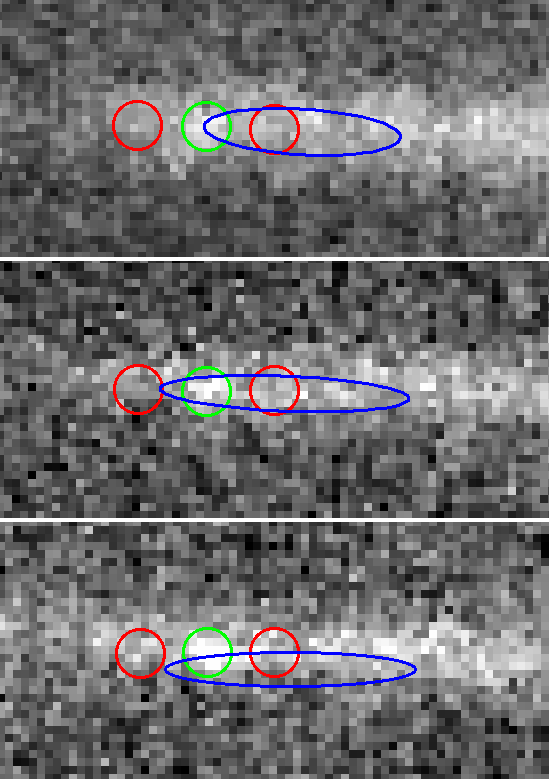}}
  \caption{Zoom into Fig.~\ref{fig:vlt_composites} showing the nucleus region on 02 (top), 30 (middle), and 31 (bottom) December 2024. The position uncertainty ellipse predicted by JPL Horizons is shown in blue. Green circles show the apertures in which the nucleus flux was measured, and red circles indicate the apertures used to assess the local background.}
  \label{fig:nucleus}
\end{figure}

\begin{figure}
  \resizebox{\hsize}{!}{\includegraphics{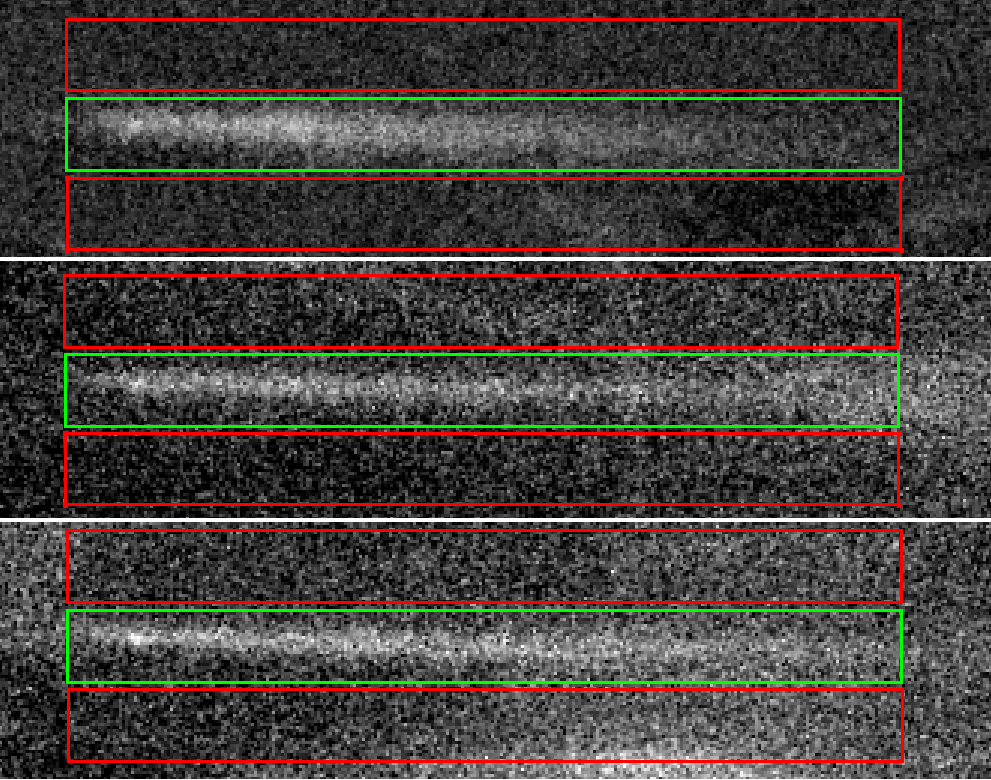}}
  \caption{Cut-out from Fig.~\ref{fig:vlt_composites} showing the regions used to measure the total dust brightness (green rectangle) and the corresponding background (red) on 02 (top), 30 (middle), and 31 (bottom) December 2024. Each region has a size of 278$\times$24 pixels or 70\arcsec$\times$6\arcsec.}
  \label{fig:tail}
\end{figure}

Figure~\ref{fig:2024_ZTF_tail} shows the stacked ZTF images from August/September 2024 (Section~\ref{sec:ZTF_2024}).
Figures~\ref{fig:nucleus} and \ref{fig:tail} illustrate the photometric apertures that were used to measure the brightness of the nucleus, dust tail, and the respective background levels in the stacked ESO images.

\twocolumn


\onecolumn
\section{Ejection perpendicular to the orbital motion}

\begin{figure}[h]
  \includegraphics[width=0.5\textwidth]{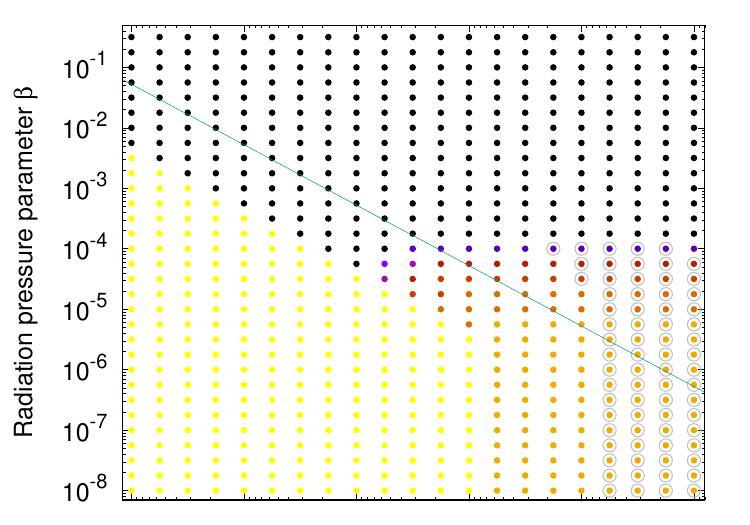}
  \includegraphics[width=0.5\textwidth]{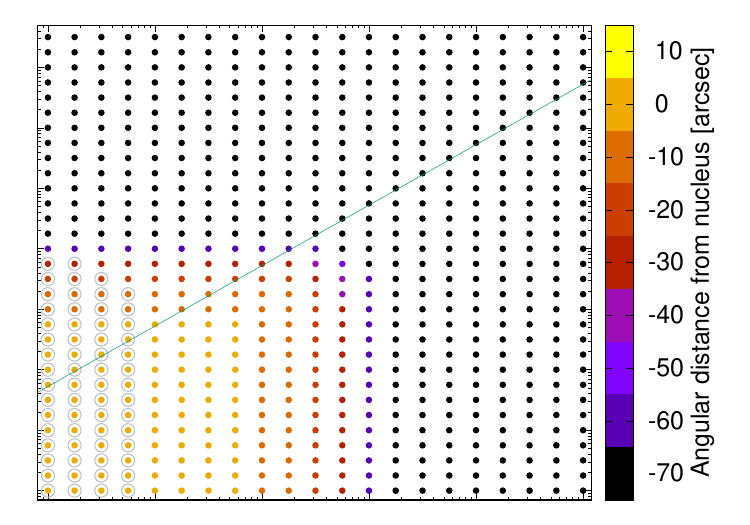}\\
   \includegraphics[width=0.5\textwidth]{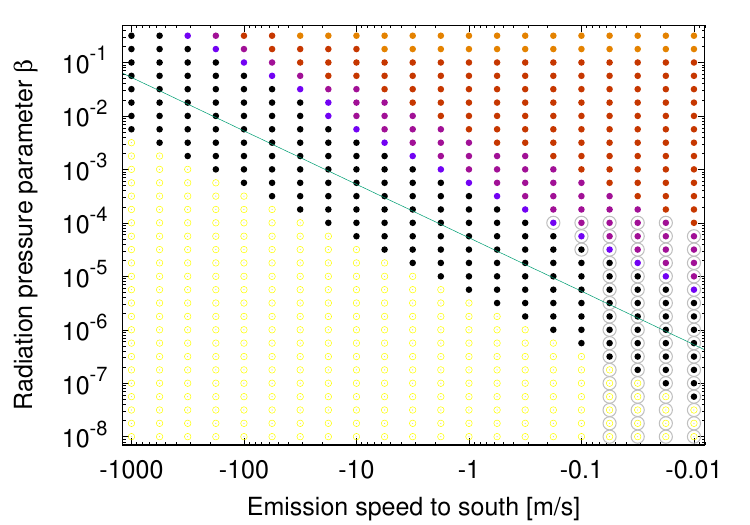}
  \includegraphics[width=0.5\textwidth]{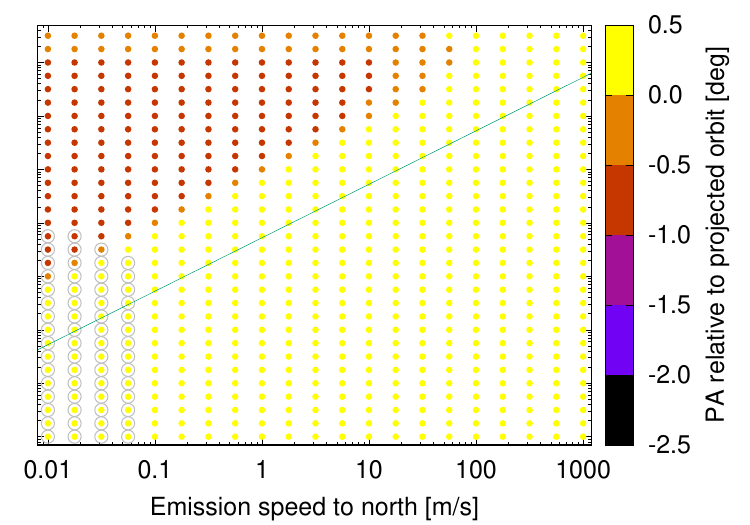}
  \caption{Same as Fig.~\ref{fig:vlt_speed_beta} for emission perpendicular to the orbital plane of JN16. Only the yellow symbols in the bottom left panel are open, indication location east of the nucleus.}
  \label{fig:vlt_speed_beta_perp}
\end{figure}

\begin{figure}[h]
  \includegraphics[width=0.5\textwidth]{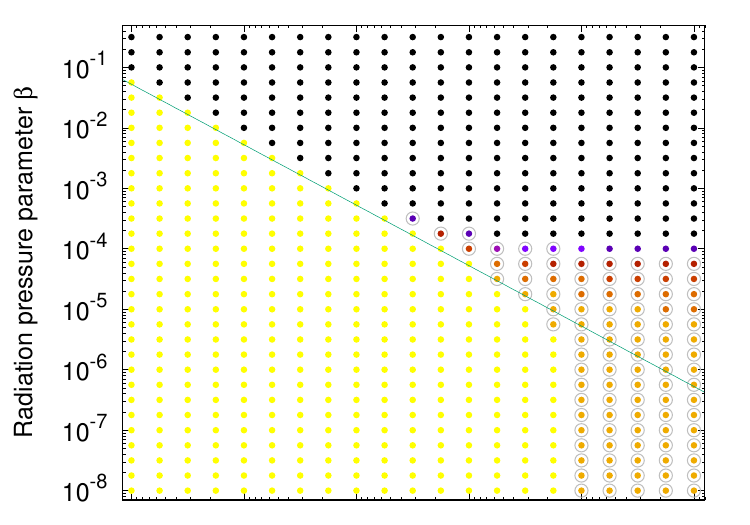}
  \includegraphics[width=0.5\textwidth]{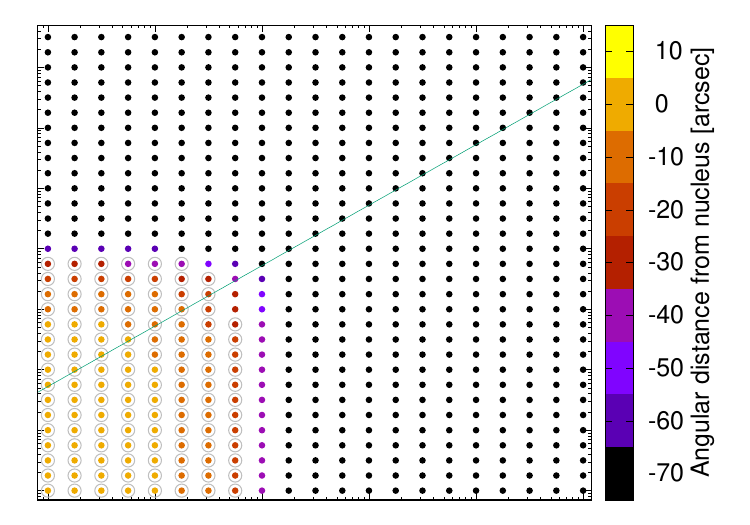}\\
   \includegraphics[width=0.5\textwidth]{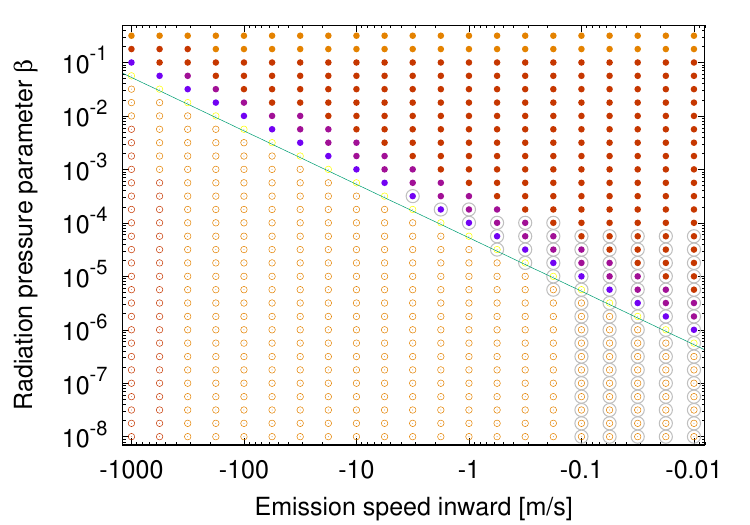}
  \includegraphics[width=0.5\textwidth]{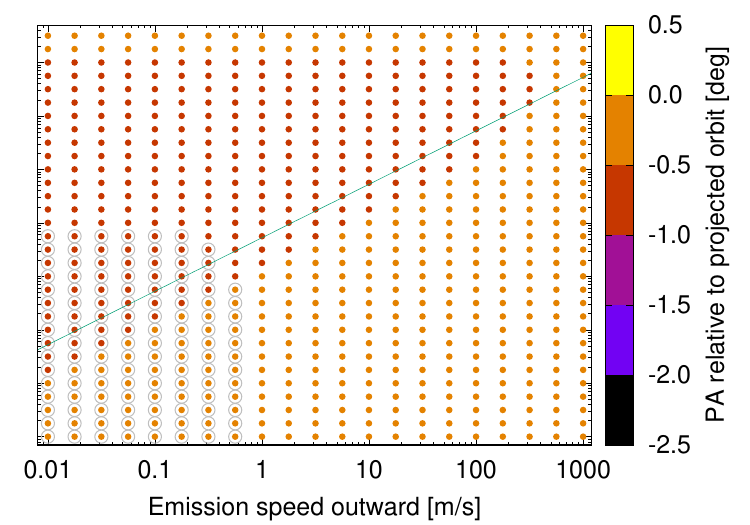}
     \caption{Same as Fig.~\ref{fig:vlt_speed_beta} for emission parallel to the orbital plane and perpendicular to the orbital motion of JN16. Only the yellow symbols in the bottom left panel are open, indicating location to the east of the nucleus.}
  \label{fig:vlt_speed_beta_rad}
\end{figure}

Figures~\ref{fig:vlt_speed_beta_perp} and \ref{fig:vlt_speed_beta_rad} describe the image locations of particles emitted to directions perpendicular to the orbital motion of JN16, forming a right-handed coordinate system with the parallel speed component investigated in Fig.~\ref{fig:vlt_speed_beta}.

Particles emitted to the north are generally projected north of the projected orbit (yellow filled symbols in Fig~\ref{fig:vlt_speed_beta_perp}, unless their speed is sufficiently small and radiation pressure is sufficiently strong that the particles are located at sufficient outwards distance from the 3d-orbit and sufficiently close to the plane that they appear projected to the south of the orbit, even through physically north of it. Particles emitted to the south are also projected to the south (because the observer was located south of the plane), either east (yellow open symbols in bottom left panel of Fig~\ref{fig:vlt_speed_beta_perp}) or west of JN16 (filled non-yellow symbols). All particles present in the visible tail must have had $|v_\perp|<$ 0.1\,m/s and $\beta<10^{-4}$.

Particles emitted in the direction roughly away from the Sun (`outwards' in Fig.~\ref{fig:vlt_speed_beta_rad}) tend to be projected close to the synchrone PA (relative PA of -0.7$^\circ$, red filled points in bottom right panel). Particles emitted inwards are projected to the north of the orbital plane and east of the nucleus (open symbols in bottom left panel), hence located inside the orbit, if the influence of radiation pressure is sufficiently small ($\beta < v$/19\,000\,m/s, below green line). Otherwise, radiation pressure compensates the inward ejection and pushes the particles to outside the orbit of JN16 and trailing it in its orbital motion (filled symbols in bottom left panel). Particles located inside the visible tail must have $\beta<10^{\-4}$ and emission speeds $<$1\,m/s.

\end{appendix}

\end{document}